\documentclass[%
 reprint,
 amsmath,amssymb,
 longbibliography,
 aps,
 pre,
]{revtex4-1}

\usepackage{calc}

\usepackage[normalem]{ulem}
\usepackage{graphicx}
\usepackage{bm}
\usepackage{xcolor}
\usepackage{hyperref}
\usepackage{amsmath}
\usepackage{amssymb}
\usepackage[utf8]{inputenc}
\usepackage{ulem}
\usepackage{xr}
\usepackage{outlines}
\usepackage{bbm}
\usepackage{mathtools}
\usepackage{graphicx}
\usepackage{hhline}
\usepackage{mathrsfs} 
\usepackage{float} %
\usepackage{tikz}
\usetikzlibrary{arrows}

\def\env{\bm{\xi}}

\def\Z{\mathbb{Z}}

\def\randMeasure{\nu}
\def\expEnv{\mathbb{E}^{\env}}

\newcommand{\expAnnealed}[1]{\mathbb{E}_{\nu}\left[ #1 \right]}

\def\varEnv{\mathrm{Var}^{\env}}
\def\varAnnealed{\mathrm{Var}_{\nu}}

\def\extCoef{\lambda_{\mathrm{ext}}}
\def\dExt{D_{\mathrm{ext}}}

\def\dExt{D_{\mathrm{ext}}}
\def\centeringTerm{c_N}
\def\driftVar{\varAnnealed\left(\expEnv[Y]\right)}
\def\localCoeff{\expAnnealed{\varEnv(Y)}}

\newcommand{\diffRandomWalk}[1]{\Delta\left( #1 \right)}
\newcommand{\absMeasure}[1]{\tilde{\mu}\left( #1 \right)}

\newcommand{\expTotal}[1]{\mathbf{E}\left[ #1 \right]}
\newcommand{\varTotal}[1]{\mathbf{Var}\left( #1 \right)}

\newcommand{\meanOmega}[1]{\expAnnealed{\xi\left(#1\right)}}
\newcommand{\invMeasure}[1]{\mu({#1})}

\def\envMax{\mathrm{Env}_t^N}
\def\max{\mathrm{Max}_t^N}
\def\samMax{\mathrm{Sam}_t^N}

\def\envFPT{\mathrm{Env}_L^N}
\def\tauL{\tau_L}
\def\min{\mathrm{Min}_L^N}
\def\samFPT{\mathrm{Sam}_L^N}

\begin{document}

\title{Extreme Diffusion Measures Statistical Fluctuations of the Environment}
\author{Jacob Hass$^*$, Hindy Drillick$^\dagger$, Ivan Corwin$^\dagger$, Eric Corwin$^*$}
\affiliation{$^*$Department of Physics and Materials Science Institute, University of Oregon, Eugene, Oregon 97403, USA. \\ $^\dagger$Department of Mathematics, Columbia University, New York, New York 10027, USA.}
\date{\today}

\begin{abstract}
We consider many-particle diffusion in one spatial dimension modeled as \emph{Random Walks in a Random Environment} (RWRE). A shared short-range space-time random environment determines the jump distributions that drive the motion of the particles. We determine universal power-laws for the environment's contribution to the variance of the extreme first passage time and extreme location. We show that the prefactors rely upon a single \emph{extreme diffusion coefficient} that is equal to the ensemble variance of the local drift imposed on particles by the random environment. This coefficient should be contrasted with the Einstein diffusion coefficient, which determines the prefactor in the power-law describing the variance of a single diffusing particle and is equal to the jump variance in the ensemble averaged random environment. Thus a measurement of the behavior of extremes in many-particle diffusion yields an otherwise difficult to measure statistical property of the fluctuations of the generally hidden environment in which that diffusion occurs. We verify our theory and the universal behavior numerically over many RWRE models and system sizes.

\end{abstract}

\maketitle



\emph{Introduction.}
Beneath the still surface of a glass of water lies an invisible roiling chaos; thermal motion moves the fluid environment at every time and length scale \cite{brownXXVIIBriefAccount1828a, perrinMouvementBrownienMolecules1910, perrinMouvementBrownienMolecules1910, vonsmoluchowskiNotizUiberBerechnung1915a,  vonsmoluchowskiZurKinetischenTheorie1906a, langevinTheorieMouvementBrownien1908a, einsteinUberMolekularkinetischenTheorie1905a, einsteinZurTheorieBrownschen1906a, einsteinTheoretischeBemerkungenUber1907a, sutherlandViscosityGasesMolecular1893a, sutherlandLXXVDynamicalTheory1905a}. It is only through the motion of tracer particles that this molecular storm is rendered visible~\cite{brownXXVIIBriefAccount1828a, perrinMouvementBrownienMolecules1910, perrinMouvementBrownienRealite1909}.
The coefficient of the mean-squared displacement power-law for a single tracer particle yields a measurement of the Einstein diffusion coefficient, $D$. Classical diffusion theory~\cite{einsteinZurTheorieBrownschen1906a, einsteinUberMolekularkinetischenTheorie1905a, einsteinTheoretischeBemerkungenUber1907a, langevinTheorieMouvementBrownien1908a, sutherlandViscosityGasesMolecular1893a, sutherlandLXXVDynamicalTheory1905a, vonsmoluchowskiZurKinetischenTheorie1906a, vonsmoluchowskiNotizUiberBerechnung1915a} relates this to a microscopic statistic of the environment, namely the square of the mean-free path length divided by the mean collision time. This is far from a complete characterization of the thermal motion within the environment. The introduction of multiple tracer particles offers the possibility of gleaning further statistical information. However, classical diffusion theory stymies such an effort since it treats each particle as independent, thus replacing the richness and chaos of the environment with an ensemble average in which particles are randomly kicked in the same manner and at the same rate everywhere in the system. This erasure is a lie, but one which works well to predict the bulk or typical behavior of many tracer particles. In this paper, we show that at the edges of the bulk, the truth is exposed: the fluctuations of the extremes of many tracer particles are highly sensitive to the disorder of the environment and thus reveal a more complete statistical description of the hidden environment.

\begin{figure}[h!] 
    \centering
    \includegraphics[width=\columnwidth]{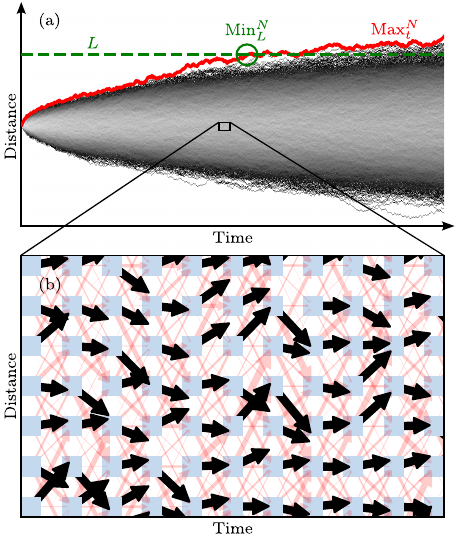}
    \caption{(a) Space-time trajectories of $N=10^5$ particles evolving in a random environment with a uniform jump distribution on the interval $\{-2,\ldots, 2\}$. The solid red line denotes the extreme location $\max$, and the green circle denotes the extreme first passage time $\min$. (b) The random environment driving the evolution in (a): blue boxes are sites, the width of the red arrows shows the probability of jumping between sites and black arrows are the average drift from a site.}
    \label{fig:Occupation}
\end{figure}

Here, we study the statistical behavior of the extreme first passage time past a barrier at location $L$ and of the extreme location at time $t$ for a system of $N$ tracer particles in a general class of \emph{random walk in random environment} (RWRE) models, see Fig. \ref{fig:Occupation}a. These models capture the fact that in real many-particle diffusion, all particles are subject to common and effectively random forces from the thermal fluctuations of the fluid environment. We show that (1) the environment's contribution to the variances of these two observables follow robust power-laws whose exponents are independent of the choice of environment; (2) the coefficient in these power-laws, as well as the time or location of onset of the power-laws, depends on the Einstein diffusion coefficient, $D$, as well another parameter $\dExt$ that is equal to the ensemble variance of the local drift imposed upon particles by the environment (see Fig. \ref{fig:Occupation}b, where $\dExt$ records the variance of the black arrows). We call $\dExt$ the \emph{extreme diffusion coefficient} as it relates the extreme behavior of many-particle diffusion to a microscopic statistic of the environment. In the RWRE model, the Einstein diffusion coefficient $D$ is related to a different microscopic statistic, namely the jump variance in the ensemble averaged environment. Thus, these two diffusion coefficients---that of Einstein which is observable from a single tracer, and that introduced here which is observable from the extremes of many tracers---offer a refined lens relative to classical diffusion theory through which to measure the statistics of the hidden environment.

\emph{RWRE models.} In place of the independent random walk model of classical diffusion theory, we consider here lattice RWRE models. These play the role of a coarse-grained continuum environment in which particles are chaotically or thermally-randomly biased in their motion by an environment quickly mixing in space and time. As such, we focus on RWRE models whose randomness is short-range correlated in time, see also~\cite{richardsonAtmosphericDiffusionShown1926a, hentschelRelativeDiffusionTurbulent1984a, bouchaudDiffusionLocalizationWaves1990a, chertkovAnomalousScalingExponents1996a, bernardAnomalousScalingNPoint1996a, jullienRichardsonPairDispersion1999a, balkovskyIntermittentDistributionInertial2001a}. This is in contrast to long-range correlated or quenched in time randomness, e.g. as in~\cite{hughesRandomWalksRandom1995b,bouchaudAnomalousDiffusionDisordered1990a, kestenLimitLawRandom1975a, sinaiLimitingBehaviorOneDimensional1983a, bouchaudClassicalDiffusionParticle1990a, burlatskyTransientRelaxationCharged1998a, ledoussalRandomWalkersOneDimensional1999a}.

To define our RWRE models, we first describe how we specify the environment, and then we describe how random walks evolve therein. Each RWRE model is specified by $\randMeasure$, a choice of probability distribution on the space of probability distributions on $\Z$. There are many ways to produce a random probability distribution on $\Z$. Perhaps the simplest involves choosing a uniform random variable on $[0,1]$ and then assigning that probability to $+1$ and its complementary probability to $-1$ (and $0$ probability assigned to all other values of $\Z$). This \emph{nearest-neighbor} model is a special case of the beta RWRE (where the uniform is replaced by a general beta distribution) introduced and studied recently in~\cite{barraquandRandomWalkBetadistributedRandom2017b, ledoussalDiffusionTimeDependentRandom2017c,barraquandModerateDeviationsDiffusion2020c, hassAnomalousFluctuationsExtremes2023b, hassFirstpassageTimeManyparticle2024,dasKPZEquationLimit2023b}. Though our theoretical results apply more generally, for our numerical simulations we will consider a few other specific non-nearest neighbor examples of $\nu$, namely the \emph{Dirichlet}, \emph{normalized i.i.d.} and \emph{random delta} distributions described in~\cite{SeeSupplementalMateriala}. For further examples, see also~\cite{parekhInvariancePrincipleKPZ2024a}.

Given $\nu$, we define the random environment as
$\env\coloneqq\big(\xi_{t,x}: x \in \mathbb{Z}, t \in \mathbb{Z}_{\geq 0} \big)$ where each $\xi_{t,x}$ is a probability distribution on $\Z$ that is sampled independently according to $\randMeasure$. The environment, $\env$, should be thought of as one instance of an environment in which several particles can diffuse from position $x$ and time $t$ using the transition probabilities $\xi_{t,x}$, i.e. the red arrows in Fig. \ref{fig:Occupation}b. We denote the product measure on the environment $\env$ by $\mathbb{P}_{\nu}$ and let $\expAnnealed{\bullet}$ and $\varAnnealed(\bullet)$ denote the associated expectation and variance. We restrict our attention to models $\randMeasure$ that produce net drift-free systems, i.e. such that $\sum_{j\in \Z} \expAnnealed{\xi(j)} j = 0$ where $\xi$ is sampled according to $\randMeasure$, and those with finite range, i.e. such that there exists an $M>0$ so that $\xi(j)=0$ if $|j|>M$. After going to a suitable moving frame, similar results to ours hold even when there is a net drift, i.e. non-zero expected mean.

Given a sample $\env$ of the environment, we define a probability measure $\mathbb{P}^{\env}$ on an arbitrary number of independent and identically distributed (i.i.d.) random walks $R^1,R^2,\ldots$ evolving in that environment, and let $\expEnv[\bullet]$ and $\varEnv(\bullet)$ denote the associated expectation and variance. Each walk starts at $R^i(0)=0$, where $R^i(t) \in \Z$ denotes its position at time $t$.  The probability that a walk at $x$ at time $t$ transitions to $x + j$ at time $t+1$ is
\begin{equation*}
    \mathbb{P}^{\env}\big(R^i(t+1) = x + j \mid R^i(t)=x\big) = \xi_{t,x}(j).
\end{equation*} 
All random walks evolve independently given $\env$ though notably walkers at the same location $x$ at the same time $t$ are subject to the same jump distribution, $\xi_{t,x}$. We define the transition probabilities given $\env$ by $p^{\env}(x,t) = \mathbb{P}^{\env}(R(t)=x)$ (we drop the superscript $i$ here and elsewhere below since each $R^i(t)$ is i.i.d.); this satisfies $p^{\env}(x,0)=\mathbf{1}_{x=0}$ and the master equation
\begin{equation}\label{eq:RecussionRelation}
    p^{\env}(x, t+1) = \sum_{j \in \Z}p^{\env}(x-j, t) \xi_{t,x-j}(j).
\end{equation}

The measure $\mathbb{P}_{\nu}$ is on the environment $\env$ and $\mathbb{P}^{\env}$ is on independent random walks given the environment $\env$. It is also useful to define $\mathbf{P}$ by $\mathbf{P}(\bullet) = \expAnnealed{\mathbb{P}^{\env}(\bullet)}$ for any event $\bullet$ defined in terms of the random walks $R^1,R^2,\ldots$, and to let $\mathbf{E}[\bullet]$ and $\mathbf{Var}[\bullet]$ denote the associated expectation and variance.  
This is the marginal distribution on many-particle diffusion trajectories that is relevant to repeated experimental studies of many-particle diffusion; $\mathbf{P}$ represents the histogram over many experimental samples of the many-particle diffusion trajectories.


Given a random walk $R^i$, we denote its first passage time past $L>0$ by $\tau_L^i$. The \emph{extreme first passage time} of $N$ random walkers past location $L$ is defined as
\begin{equation*}
\min \coloneqq \mathrm{min}(\{\tau_L^1, \ldots, \tau_L^N\}).
\end{equation*}
Given an environment $\env$, the walks $R^1,\ldots, R^N$ are i.i.d under the measure $\mathbb{P}^{\env}$, and thus so are the $\tau_L^i$. Therefore,
\begin{equation}\label{eq:MinDef}
    \mathbb{P}^{\env}(\min \leq t) = 1 - \big(1-\mathbb{P}^{\env}(\tau_L \leq t)\big)^N.
\end{equation}
We also study the \emph{extreme location} of $N$ walks at time $t$,
\begin{equation*}
    \max \coloneqq \mathrm{max}\left(\{ R^1(t), \ldots R^N(t)\}\right).
\end{equation*}
Given $\env$, under the measure $\mathbb{P}^{\env}$ on $R^1,\ldots, R^N$,
\begin{equation}\label{eq:MaxDef}
    \mathbb{P}^{\env}(\max \geq x) = 1-\big(1-\mathbb{P}^{\env}(R(t)\geq x)\big)^N.
\end{equation}

We characterize the statistics of $\min$ and $\max$ under the measure $\mathbf{P}$, i.e., when the environment is hidden as in real diffusive systems. There are two levels of randomness: first, the randomness due to the environment, $\envMax$, and second, the randomness of sampling walkers in that environment, $\samFPT$. Specifically, we define $\envFPT$ as the minimum time $t$ such that $\mathbb{P}^{\env}(\tau_L \leq t) \geq \frac{1}{N}$ and $\samFPT$ as the residual, $\samFPT \coloneqq \min - \envFPT$.  Note that $\envFPT$ only depends on the environment $\env$ and by \eqref{eq:MinDef}, $\envFPT$ satisfies $\mathbb{P}^{\env}(\min \leq \envFPT)\approx 1-e^{-1}$. Thus, $\envFPT$ captures an environment-dependent centering of $\min$ while $\samFPT$ captures sampling fluctuations around that. We similarly define $\envMax$ as the maximum position $x$ satisfying $\mathbb{P}^{\env}(R(t) \geq x) \geq \frac{1}{N}$ and $\samMax \coloneqq \max - \envMax$.



\emph{Theoretical Results.} 
The Einstein diffusion coefficient,
\begin{equation}\label{eq:edc}
D := \frac{1}{2}\sum_{j \in \Z} \expAnnealed{\xi(j)} j^2,
\end{equation}
for the RWRE is defined as the variance of the ensemble averaged jump distribution (recall that we have assumed a net drift-free system, i.e. $\sum_{j\in \Z} \expAnnealed{\xi(j)} j = 0$). With this, we find that
\begin{equation*}
    \mathbf{E}\left[\min\right] \approx \frac{L^2}{4D\ln(N)}, \quad \mathbf{E}[\max] \approx \sqrt{4 D \ln(N) t}
\end{equation*}
match the classical theory of diffusion, i.e., where the RWRE is replaced by independent random walks with Einstein diffusion coefficient $D$, see, e.g.~\cite{schussRedundancyPrincipleRole2019a, linnExtremeHittingProbabilities2022a, lawleyDistributionExtremeFirst2020a, lawleyUniversalFormulaExtreme2020a, madridCompetitionSlowFast2020a, basnayakeAsymptoticFormulasExtreme2019a}.

The variance of $\min$ and $\max$ reveals more interesting behavior. The environmental and sampling contributions to $\min$ and $\max$ are roughly independent: 
\begin{align}\label{eq:AdditionLaw1}   
\mathbf{Var}\left(\min\right) &\approx \mathbf{Var}\left(\envFPT\right) + \mathbf{Var}\left(\samFPT\right),\\
\mathbf{Var}\left(\max\right) &\approx \mathbf{Var}\left(\envMax\right) + \mathbf{Var}\left(\samMax\right).\label{eq:AdditionLaw2}   
\end{align}

The sampling fluctuations are centered Gumbel with
\begin{align*}
    \mathbf{Var}(\samFPT) \approx \frac{\pi^2 L^4}{96 D^2 \log(N)^4},\quad
    \mathbf{Var}(\samMax) \approx \frac{\pi^2 D t}{6 \ln(N)},
\end{align*}
in agreement with classical diffusion theory~\cite{schussRedundancyPrincipleRole2019a, linnExtremeHittingProbabilities2022a, lawleyDistributionExtremeFirst2020a, lawleyUniversalFormulaExtreme2020a, madridCompetitionSlowFast2020a, basnayakeAsymptoticFormulasExtreme2019a}. The environmental fluctuations follow anomalous power-laws
\begin{align}
\label{eq:envMinAsym} \mathbf{Var}(\envFPT) &\approx \extCoef \frac{\sqrt{2 \pi} L^3}{8 D^2 \log(N)^{5/2}} \\
\label{eq:envMax} \mathbf{Var}(\envMax) &\approx \extCoef \sqrt{2 \pi Dt}.
\end{align}
when
$
L \gg \extCoef \ln(N)^{3/2}$ or $t \gg \frac{\extCoef^2}{D}\ln(N)^2$, and 
\begin{equation}\label{eq:lambdaSimplified}
    \extCoef = \frac{1}{2} \frac{\dExt}{(D - \dExt)}=\frac{1}{2} \frac{\driftVar}{\localCoeff}
\end{equation}
where $D$ is the Einstein diffusion coefficient and $\dExt$ is the \emph{extreme diffusion coefficient}:
\begin{equation}\label{eq:dExt}
    \dExt := \frac{1}{2} \driftVar=  \frac{1}{2}\expAnnealed{\big( \sum_{j\in\Z} \xi(j) j \big)^2}.
\end{equation}
%

Here and below, $Y$ is a random jump distributed according to $\xi$. Thus $\dExt$ is the variance over the random environment of the drift of a single jump, i.e. how much the black arrows fluctuate over space and time in Fig \ref{fig:Occupation}b. Then $\extCoef$ is the ratio of that to the mean over the random environment of the variance of a single jump. The formula in \eqref{eq:lambdaSimplified} is derived in~\cite{SeeSupplementalMateriala} under the assumptions
\begin{equation}\label{eq.ass}
\expAnnealed{\xi(i) \xi(j)} = c \expAnnealed{\xi(i)} \expAnnealed{\xi(j)}\quad \textrm{for all } i\neq j
\end{equation}
with some fixed $c\in (0,1)$, 
and the RWRE is aperiodic (i.e., it does not live on a strict space-time sublattice). In terms of $c$, $\extCoef = \frac{1-c}{2c}$. When \eqref{eq.ass} fails, there is a more involved formula for $\extCoef$, see~\cite{parekhInvariancePrincipleKPZ2024a,hassUniversalKPZFluctuations2024,SeeSupplementalMateriala}. When aperiodicity fails, e.g. for the nearest-neighbor RWRE, an additional factor arises in \eqref{eq:lambdaSimplified}, see \cite{SeeSupplementalMateriala}. 

The extreme diffusion coefficient necessarily satisfies $\dExt\in [0,D]$. When $\dExt = D$ (hence $\extCoef = \infty$), $\xi$ is supported at a single, yet random site, hence a perfectly sticky environment of coalescing random walkers. There is a scaling limit where $\dExt \rightarrow D$ as time is rescaled which leads to sticky Brownian motions as studied, for instance, in~\cite{dasKPZEquationLimit2023b, parekhInvariancePrincipleKPZ2024a}. When $\dExt = 0$, (hence $\extCoef= 0$), the drift $\mathbb{E}^{\env}[Y]$ becomes deterministic, thus with probability $1$ under $\mathbb{P}_{\nu}$,  $\sum_{j} \xi(j) j$ is constant, and (by the net drift-free assumption) equal to $0$, see~\cite{parekhInvariancePrincipleKPZ2024a, hassSuperUniversalBehaviorExtreme2024a}.

\begin{figure*}[ht!]
\hspace{-2cm}
\begin{minipage}{.4\textwidth}
\includegraphics[scale=0.6]{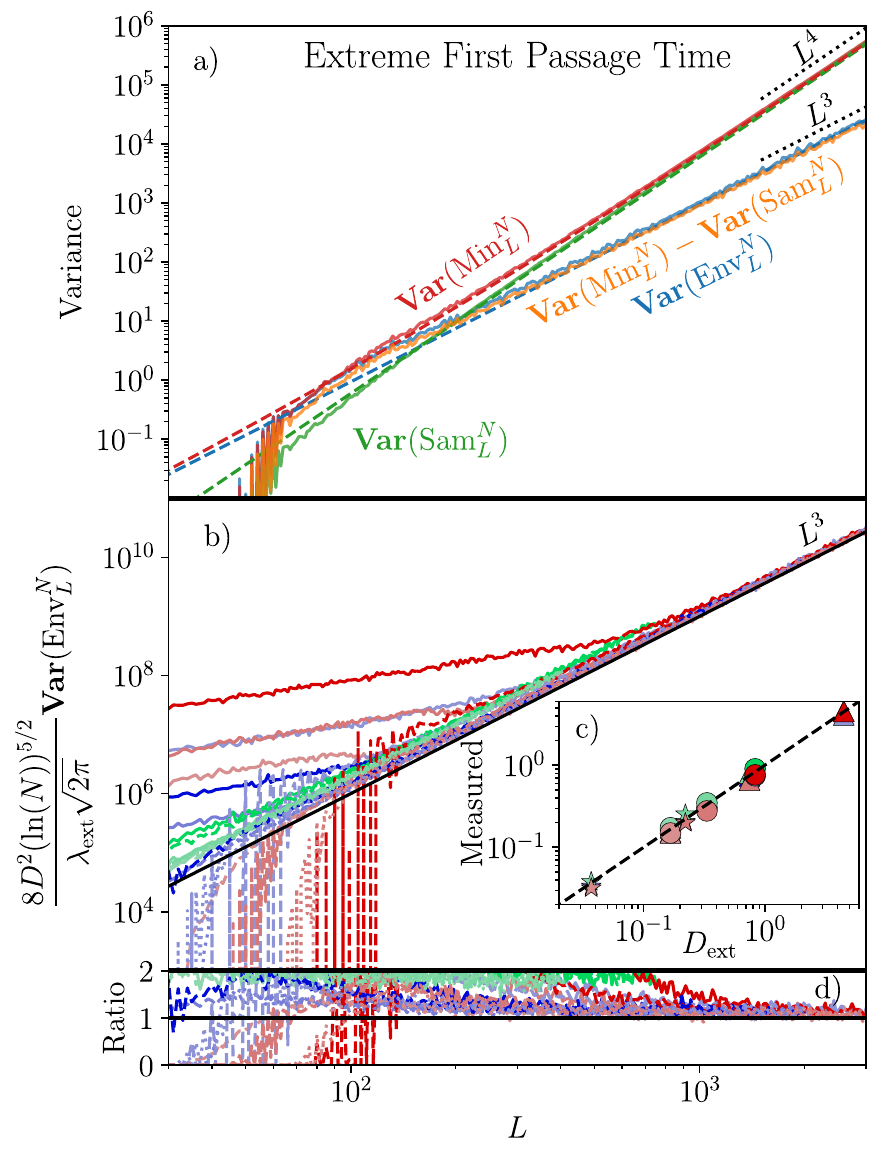}
\end{minipage}\qquad\qquad\qquad
\begin{minipage}{.4\textwidth}
\includegraphics[scale=0.6]{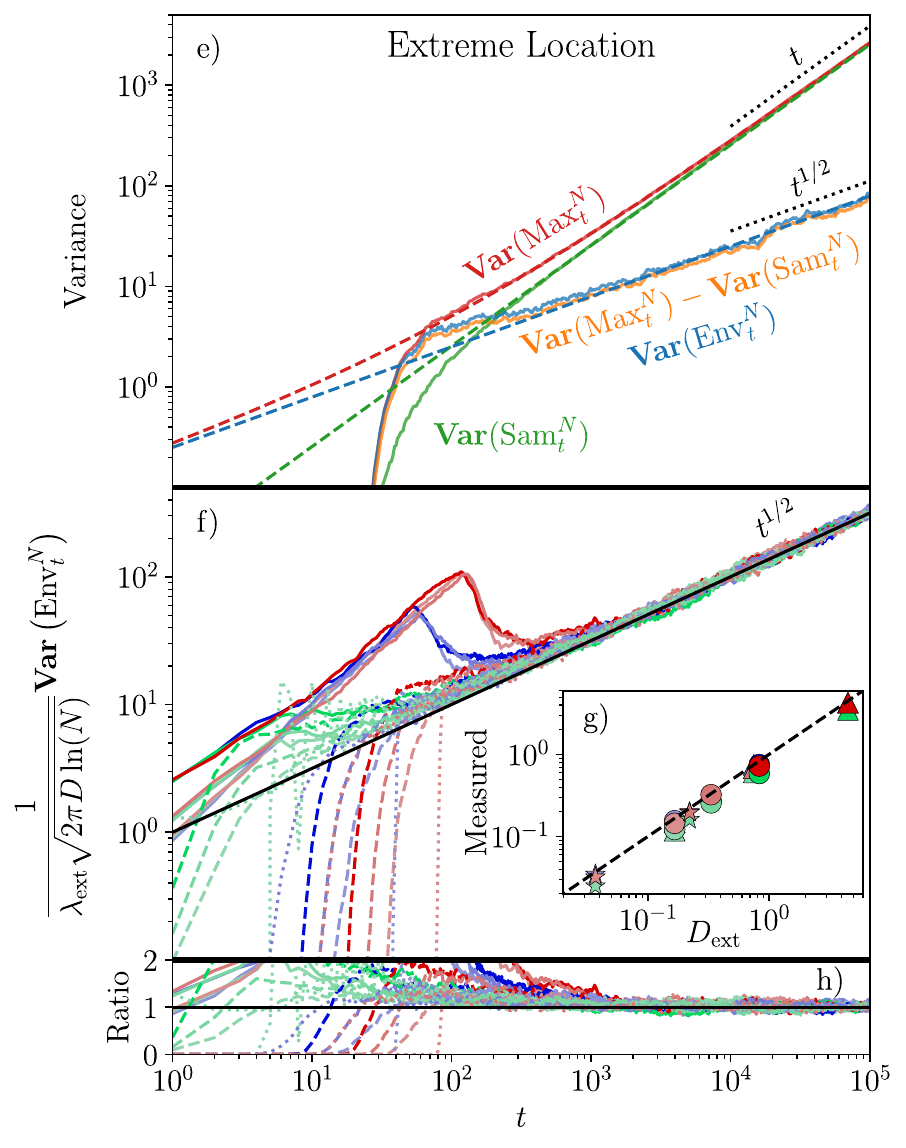}
\end{minipage}
\caption{Plot of the measured variances for the extreme first passage time (a) and extreme location (e) for systems of $N=10^{28}$ and the uniform distribution with $D=1$. b,f) Collapse of the environmental variance as a function of $L$ and $t$, respectively. $N=10^2, 10^{14}$ and $10^{28}$ is plotted in red, blue and green respectively; the uniform, Dirichlet and random delta distributions for $\randMeasure$ (see \cite{SeeSupplementalMateriala} for definitions) are rendered as dashed, dotted and solid lines respectively; the color saturation is proportional to the Einstein diffusion coefficient $D$. Inset, c,g) Plot of the measured $\dExt$ against the true value of $\dExt$. The dashed line represents equality. The uniform, Dirichlet and random delta distributions are labeled with a circle, star and triangle, respectively. d,h) Plot of ratios of measured environmental variance vs. theoretical prediction in \eqref{eq:envMinAsym} and \eqref{eq:envMax} as a function of $L$ and $t$, respectively.}
  \label{fig:FPTMax}
\end{figure*}

\emph{Theoretical Methods.}
Our derivation~\cite{SeeSupplementalMateriala} of the extreme first passage and location statistics closely follows~\cite{hassAnomalousFluctuationsExtremes2023b, hassFirstpassageTimeManyparticle2024} and relies on moderate deviation asymptotics for the tail probability $\mathbb{P}^{\env}(R(t) \geq x)$ with $x\propto t^{3/4}$, where $\env$ is distributed according to $\mathbb{P}_{\nu}$. We rely on results from~\cite{parekhInvariancePrincipleKPZ2024a,hassUniversalKPZFluctuations2024} that prove convergence in this regime of the tail probability to the KPZ equation~\cite{kardarDynamicScalingGrowing1986a, corwinKardarParisiZhang2012a}. The occurrence of the KPZ equation under moderate deviation scaling was predicted (without a formula for the KPZ coefficients) first in \cite{ledoussalDiffusionTimeDependentRandom2017c} and confirmed with coefficients for the beta RWRE in \cite{barraquandModerateDeviationsDiffusion2020c} and general nearest-neighbor RWRE in \cite{dasKPZEquationLimit2023c}, all prior to the general case addressed in \cite{parekhInvariancePrincipleKPZ2024a,hassUniversalKPZFluctuations2024}.
The KPZ convergence results imply 
\begin{align*}
 \varTotal{\envFPT} &\approx \frac{L^4}{(4D)^2 \ln(N)^4} \mathrm{Var}\left(h\left(\frac{32 \extCoef^2 \ln(N)^3}{L^2}, 0 \right)\right),\\
 \varTotal{\envMax} &\approx \frac{D t}{\ln(N)} \mathrm{Var}\left( h \left( \frac{8 \extCoef^2 \ln(N)^2}{D t}, 0 \right) \right),
\end{align*}
where $h(t,x)$ is the narrow-wedge solution to the KPZ equation at time $t$ and location $x$. The power-laws and their regime of applicability come from the short-time Edwards-Wilkinson asymptotics of the KPZ equation, see e.g.~\cite{amirProbabilityDistributionFree2011a,calabreseFreeEnergyDistributionDirected2010a}. The formulas for $\extCoef$ from~\cite{parekhInvariancePrincipleKPZ2024a,hassUniversalKPZFluctuations2024} are considerably more complicated than \eqref{eq:lambdaSimplified} and hold for general RWRE, even those in violation of \eqref{eq.ass} or the aperiodicity assumption. In particular, $\extCoef$ involves the two-point motion, i.e. the law of $R^1$ and $R^2$ under $\mathbf{P}$, via the expected change in its difference $|R^1(t)-R^2(t)|$ in a unit of time when initialized under its invariant measure. That expression arises from comparing the pair intersection time for the two-point motion (as arises in the second moment formula for the tail probability) with the limiting Brownian intersection local time (as relates to the exponentiated KPZ equation via the replica method \cite{kardarReplicaBetheAnsatz1987, bertiniStochasticHeatEquation1995a}) via a discrete version of the Tanaka formula.

\emph{Numerical Methods.} As in~\cite{hassAnomalousFluctuationsExtremes2023b,hassFirstpassageTimeManyparticle2024} (see also~\cite{SeeSupplementalMateriala}), for each environment $\env$ sampled according to $\nu$ we numerically compute the total probability mass absorbed up to time $t$, $\mathbb{P}^{\env}(\tau_L \leq t)$, by solving \eqref{eq:RecussionRelation} with an absorbing boundary condition at $L$.  From this distribution we compute $\envFPT$ as well as the distribution of $\min$ using equation \eqref{eq:MinDef}. Finally, we compute the distribution for $\samFPT$.  Likewise, we numerically compute the total probability mass at a location $x$ and time $t$, $\mathbb{P}^{\env}(R(t) = x)$, by solving \eqref{eq:RecussionRelation}.  From this we compute $\envMax$, $\max$, and $\samMax$.
All reported quantities are measured by simulating 500 different systems for a given $\nu$ and $N$.

The extreme diffusion coefficient, $\dExt$, can be independently measured from both the extreme first passage and extreme location statistics. Using  \eqref{eq:lambdaSimplified} we find $\dExt = \frac{2\extCoef D}{1 + 2 \extCoef}$ ($D$ is computed using \eqref{eq:edc} though it could also be recovered numerically). For the extreme first passage time, $\extCoef$ is computed as 
$
    \frac{8 D^2 \log(N)^{5/2}} {\sqrt{2 \pi} L^3} \left( \varTotal{\min} - \varTotal{\samFPT} \right),
$
whereas for the extreme location, $\extCoef$ is computed as
$
    \frac{1}{\sqrt{2 \pi D t}} \left( \varTotal{\max} - \varTotal{\samMax} \right)$.


\emph{Numerical Results.} 
Figs. \ref{fig:FPTMax}a,e show that our theoretical prediction for each relevant variance is asymptotically accurate and that the addition laws in \eqref{eq:AdditionLaw1} and \eqref{eq:AdditionLaw2} are justified. Figs \ref{fig:FPTMax}b,f show the collapse of the scaled environmental variance for several choices of $\nu$. Although our predictions assume $N$ is large, they are accurate for systems as small as $N=100$. This is also shown in Figs. \ref{fig:FPTMax}d,h since the ratio of the measured environmental variance to the theoretical prediction goes to 1, asymptotically. Figs. \ref{fig:FPTMax}c,g show that our numerically computed values of $\dExt$ from both kinds of measurements match the theoretical value, falling onto the line of equality. Since $\varTotal{\samFPT}$, and similarly $\varTotal{\samMax}$, rely only on $D$, a measurement of $\min$, or $\max$ can be translated to a measurement of $\dExt$.

\emph{Conclusion.}
We have demonstrated theoretically and confirmed numerically (see Fig. \ref{fig:FPTMax}) that the environmental contributions to the variance of the extreme first passage time and variance of the extreme location of many tracer particles in a random environment display anomalous power-laws \eqref{eq:envMinAsym} and \eqref{eq:envMax} with powers that are universal with respect to the details of the environment. The coefficients in these power-laws are determined by the Einstein diffusion coefficient and a new extreme diffusion coefficient $\dExt$ that we defined here in \eqref{eq:dExt} as the ensemble variance of the drift imposed on particles by the environment. We numerically confirm the addition laws \eqref{eq:AdditionLaw1} \eqref{eq:AdditionLaw2} that allow us to recover the environmental variance power-law and coefficient by measuring the variance of the extreme first passage time and location, along with the variance due to sampling (which itself is asymptotically determined only by the Einstein diffusion coefficient). Thus, the extreme diffusion coefficient, a quantity inherent to the hidden environment and not readily measurable, can potentially be measured by focusing on the measurable extreme behavior of many tracer particles. While we have focused herein on lattice models, the theory of extreme diffusion should extend to systems with continuous space and time dimensions provided sufficiently fast mixing of the random environment in both dimensions (see related work \cite{brockingtonEdgeCloudBrownian2022a}). There are several outstanding theoretical challenges including 1) establishing the role of correlation length and time scales in defining the continuum extreme diffusion coefficient, 2) extending extreme diffusion theory to higher spatial dimensions, 3) understanding whether there exists a fluctuation-dissipation type relation for the extreme diffusion coefficient, and if so, what is the relevant notion of extreme drag, and 4) testing extreme diffusion in experimental systems.

\emph{Acknowledgements. } This work was funded by the W.M. Keck Foundation Science and Engineering grant on “Extreme Diffusion”. I.C. also wishes to acknowledge ongoing support from  NSF DMS:1811143, DMS:1937254, and DMS:2246576, and the Simons Foundation through an Investigator Grant (Award ID 929852). E.I.C. wishes to acknowledge ongoing support from the Simons Foundation for the collaboration Cracking the Glass Problem via Award No. 454939. H.D. wishes to acknowledge ongoing support from the NSF GRFP DGE:2036197. We thank Shalin Parekh for many helpful discussions. This work benefited from access to the University of Oregon high performance computing cluster, Talapas. 

\bibliographystyle{unsrtnat} 
\bibliography{main}

\begin{thebibliography}{50}
\providecommand{\natexlab}[1]{#1}
\providecommand{\url}[1]{\texttt{#1}}
\expandafter\ifx\csname urlstyle\endcsname\relax
  \providecommand{\doi}[1]{doi: #1}\else
  \providecommand{\doi}{doi: \begingroup \urlstyle{rm}\Url}\fi

\bibitem[Brown(1828)]{brownXXVIIBriefAccount1828a}
Robert Brown.
\newblock {{XXVII}}. {{A}} brief account of microscopical observations made in
  the months of {{June}}, {{July}} and {{August}} 1827, on the particles
  contained in the pollen of plants; and on the general existence of active
  molecules in organic and inorganic bodies.
\newblock \emph{The philosophical magazine}, 4\penalty0 (21):\penalty0
  161--173, 1828.

\bibitem[Perrin(1910)]{perrinMouvementBrownienMolecules1910}
Jean Perrin.
\newblock Mouvement brownien et mol{\'e}cules.
\newblock \emph{JCP}, 8:\penalty0 57--91, 1910.

\bibitem[Von~Smoluchowski(1915)]{vonsmoluchowskiNotizUiberBerechnung1915a}
M.~Von~Smoluchowski.
\newblock Notiz {{Uiber Die Berechnung Der Brownschen Molekularbewegung Bei Der
  Ehrenhaft-Millikanschen Versuchsanordning}}.
\newblock \emph{Phys. Z}, 16:\penalty0 318--321, 1915.

\bibitem[{von Smoluchowski}(1906)]{vonsmoluchowskiZurKinetischenTheorie1906a}
M.~{von Smoluchowski}.
\newblock Zur {{Kinetischen Theorie Der Brownschen Molekularbewegung Und Der
  Suspensionen}}.
\newblock \emph{Annalen der Physik}, 326\penalty0 (14):\penalty0 756--780,
  1906.
\newblock \doi{10.1002/andp.19063261405}.

\bibitem[Langevin(1908)]{langevinTheorieMouvementBrownien1908a}
Paul Langevin.
\newblock Sur {{La Th{\'e}orie Du Mouvement Brownien}}.
\newblock 146:\penalty0 530--533, 1908.

\bibitem[Einstein(1905)]{einsteinUberMolekularkinetischenTheorie1905a}
A.~Einstein.
\newblock {\"U}ber {{Die}} von {{Der Molekularkinetischen Theorie Der W{\"a}rme
  Geforderte Bewegung}} von in {{Ruhenden Fl{\"u}ssigkeiten Suspendierten
  Teilchen}}.
\newblock \emph{Annalen der Physik}, 322\penalty0 (8):\penalty0 549--560, 1905.
\newblock ISSN 1521-3889.
\newblock \doi{10.1002/andp.19053220806}.

\bibitem[Einstein(1906)]{einsteinZurTheorieBrownschen1906a}
A.~Einstein.
\newblock Zur {{Theorie Der Brownschen Bewegung}}.
\newblock \emph{Annalen der Physik}, 324\penalty0 (2):\penalty0 371--381,
  January 1906.
\newblock ISSN 1521-3889.
\newblock \doi{10.1002/andp.19063240208}.

\bibitem[Einstein(1907)]{einsteinTheoretischeBemerkungenUber1907a}
A.~Einstein.
\newblock Theoretische {{Bemerkungen {\"U}ber Die Brownsche Bewegung}}.
\newblock \emph{Zeitschrift f{\"u}r Elektrochemie und angewandte physikalische
  Chemie}, 13\penalty0 (6):\penalty0 41--42, February 1907.
\newblock ISSN 0005-9021.
\newblock \doi{10.1002/bbpc.19070130602}.

\bibitem[Sutherland(1893)]{sutherlandViscosityGasesMolecular1893a}
William Sutherland.
\newblock The {{Viscosity}} of {{Gases}} and {{Molecular Force}}.
\newblock \emph{Philosophical Magazine Series 5}, 36\penalty0 (223):\penalty0
  507--531, 1893.
\newblock ISSN 1941-5982.
\newblock \doi{10.1080/14786449308620508}.

\bibitem[Sutherland(1905)]{sutherlandLXXVDynamicalTheory1905a}
William Sutherland.
\newblock {{LXXV}}. {{A Dynamical Theory}} of {{Diffusion}} for
  {{Non-Electrolytes}} and the {{Molecular Mass}} of {{Albumin}}.
\newblock \emph{The London, Edinburgh, and Dublin Philosophical Magazine and
  Journal of Science}, 9\penalty0 (54):\penalty0 781--785, June 1905.
\newblock \doi{10.1080/14786440509463331}.

\bibitem[Perrin(1909)]{perrinMouvementBrownienRealite1909}
Jean~Baptiste Perrin.
\newblock Le {{Mouvement Brownien}} et la {{R{\'e}alit{\'e} Moleculaire}}.
\newblock \emph{Ann. Chimi. Phys.}, 18\penalty0 (8):\penalty0 5--114, 1909.

\bibitem[Richardson and Walker(1926)]{richardsonAtmosphericDiffusionShown1926a}
Lewis~Fry Richardson and Gilbert~Thomas Walker.
\newblock Atmospheric {{Diffusion Shown}} on a {{Distance-Neighbour Graph}}.
\newblock \emph{Proceedings of the Royal Society of London. Series A,
  Containing Papers of a Mathematical and Physical Character}, 110\penalty0
  (756):\penalty0 709--737, April 1926.
\newblock \doi{10.1098/rspa.1926.0043}.

\bibitem[Hentschel and
  Procaccia(1984)]{hentschelRelativeDiffusionTurbulent1984a}
H.~G.~E. Hentschel and Itamar Procaccia.
\newblock Relative {{Diffusion}} in {{Turbulent Media}}: {{The Fractal
  Dimension}} of {{Clouds}}.
\newblock \emph{Physical Review A}, 29\penalty0 (3):\penalty0 1461--1470, March
  1984.
\newblock \doi{10.1103/PhysRevA.29.1461}.

\bibitem[Bouchaud(1990)]{bouchaudDiffusionLocalizationWaves1990a}
J.~P. Bouchaud.
\newblock Diffusion and {{Localization}} of {{Waves}} in a {{Time-Varying
  Random Environment}}.
\newblock \emph{Europhysics Letters (EPL)}, 11\penalty0 (6):\penalty0 505--510,
  March 1990.
\newblock ISSN 0295-5075.
\newblock \doi{10.1209/0295-5075/11/6/004}.

\bibitem[Chertkov and Falkovich(1996)]{chertkovAnomalousScalingExponents1996a}
M.~Chertkov and G.~Falkovich.
\newblock Anomalous {{Scaling Exponents}} of a {{White-Advected Passive
  Scalar}}.
\newblock \emph{Physical Review Letters}, 76\penalty0 (15):\penalty0
  2706--2709, April 1996.
\newblock \doi{10.1103/PhysRevLett.76.2706}.

\bibitem[Bernard et~al.(1996)Bernard, Gawedzki, and
  Kupiainen]{bernardAnomalousScalingNPoint1996a}
Denis Bernard, Krzysztof Gawedzki, and Antti Kupiainen.
\newblock Anomalous {{Scaling}} in the {{N-Point Functions}} of {{Passive
  Scalar}}.
\newblock \emph{Physical Review E}, 54\penalty0 (3):\penalty0 2564--2572,
  September 1996.
\newblock ISSN 1063-651X, 1095-3787.
\newblock \doi{10.1103/PhysRevE.54.2564}.

\bibitem[Jullien et~al.(1999)Jullien, Paret, and
  Tabeling]{jullienRichardsonPairDispersion1999a}
Marie-Caroline Jullien, J{\'e}r{\^o}me Paret, and Patrick Tabeling.
\newblock Richardson {{Pair Dispersion}} in {{Two-Dimensional Turbulence}}.
\newblock \emph{Physical Review Letters}, 82\penalty0 (14):\penalty0
  2872--2875, April 1999.
\newblock \doi{10.1103/PhysRevLett.82.2872}.

\bibitem[Balkovsky et~al.(2001)Balkovsky, Falkovich, and
  Fouxon]{balkovskyIntermittentDistributionInertial2001a}
E.~Balkovsky, G.~Falkovich, and A.~Fouxon.
\newblock Intermittent {{Distribution}} of {{Inertial Particles}} in
  {{Turbulent Flows}}.
\newblock \emph{Physical Review Letters}, 86\penalty0 (13):\penalty0
  2790--2793, March 2001.
\newblock \doi{10.1103/PhysRevLett.86.2790}.

\bibitem[Hughes(1995)]{hughesRandomWalksRandom1995b}
Barry~D. Hughes.
\newblock \emph{Random {{Walks}} and {{Random Environments}}: {{Random
  Walks}}}.
\newblock Clarendon Press, 1995.
\newblock ISBN 978-0-19-853788-5.

\bibitem[Bouchaud and Georges(1990)]{bouchaudAnomalousDiffusionDisordered1990a}
Jean-Philippe Bouchaud and Antoine Georges.
\newblock Anomalous {{Diffusion}} in {{Disordered Media}}: {{Statistical
  Mechanisms}}, {{Models}} and {{Physical Applications}}.
\newblock \emph{Physics Reports}, 195\penalty0 (4):\penalty0 127--293, November
  1990.
\newblock ISSN 0370-1573.
\newblock \doi{10.1016/0370-1573(90)90099-N}.

\bibitem[Kesten et~al.(1975)Kesten, Kozlov, and
  Spitzer]{kestenLimitLawRandom1975a}
H.~Kesten, M.~V. Kozlov, and F.~Spitzer.
\newblock A limit law for random walk in a random environment.
\newblock \emph{Compositio Mathematica}, 30\penalty0 (2):\penalty0 145--168,
  1975.

\bibitem[Sinai(1983)]{sinaiLimitingBehaviorOneDimensional1983a}
Y.~Sinai.
\newblock The {{Limiting Behavior}} of a {{One-Dimensional Random Walk}} in a
  {{Random Medium}}.
\newblock \emph{Theory of Probability \& Its Applications}, 27\penalty0
  (2):\penalty0 256--268, January 1983.
\newblock ISSN 0040-585X.
\newblock \doi{10.1137/1127028}.

\bibitem[Bouchaud et~al.(1990)Bouchaud, Comtet, Georges, and
  Le~Doussal]{bouchaudClassicalDiffusionParticle1990a}
J.~P Bouchaud, A~Comtet, A~Georges, and P~Le~Doussal.
\newblock Classical {{Diffusion}} of a {{Particle}} in a {{One-Dimensional
  Random Force Field}}.
\newblock \emph{Annals of Physics}, 201\penalty0 (2):\penalty0 285--341, August
  1990.
\newblock ISSN 0003-4916.
\newblock \doi{10.1016/0003-4916(90)90043-N}.

\bibitem[Burlatsky and Deutch(1998)]{burlatskyTransientRelaxationCharged1998a}
S.~F. Burlatsky and John~M. Deutch.
\newblock Transient {{Relaxation}} of a {{Charged Polymer Chain Subject}} to an
  {{External Field}} in a {{Random Tube}}.
\newblock \emph{The Journal of Chemical Physics}, 109\penalty0 (6):\penalty0
  2572--2578, August 1998.
\newblock ISSN 0021-9606.
\newblock \doi{10.1063/1.476831}.

\bibitem[Le~Doussal et~al.(1999)Le~Doussal, Monthus, and
  Fisher]{ledoussalRandomWalkersOneDimensional1999a}
Pierre Le~Doussal, C{\'e}cile Monthus, and Daniel~S. Fisher.
\newblock Random {{Walkers}} in {{One-Dimensional Random Environments}}:
  {{Exact Renormalization Group Analysis}}.
\newblock \emph{Physical Review E}, 59\penalty0 (5):\penalty0 4795--4840, May
  1999.
\newblock \doi{10.1103/PhysRevE.59.4795}.

\bibitem[Barraquand and
  Corwin(2017)]{barraquandRandomWalkBetadistributedRandom2017b}
Guillaume Barraquand and Ivan Corwin.
\newblock Random-{{Walk}} in {{Beta-distributed Random Environment}}.
\newblock \emph{Probability Theory and Related Fields}, 167\penalty0
  (3):\penalty0 1057--1116, April 2017.
\newblock ISSN 1432-2064.
\newblock \doi{10.1007/s00440-016-0699-z}.

\bibitem[Le~Doussal and
  Thiery(2017)]{ledoussalDiffusionTimeDependentRandom2017c}
Pierre Le~Doussal and Thimoth{\'e}e Thiery.
\newblock Diffusion in {{Time-Dependent Random Media}} and the
  {{Kardar-Parisi-Zhang Equation}}.
\newblock \emph{Physical Review E}, 96\penalty0 (1):\penalty0 010102, July
  2017.
\newblock \doi{10.1103/PhysRevE.96.010102}.

\bibitem[Barraquand and
  Doussal(2020)]{barraquandModerateDeviationsDiffusion2020c}
Guillaume Barraquand and Pierre~Le Doussal.
\newblock Moderate {{Deviations}} for {{Diffusion}} in {{Time Dependent Random
  Media}}.
\newblock \emph{Journal of Physics A: Mathematical and Theoretical},
  53\penalty0 (21):\penalty0 215002, May 2020.
\newblock ISSN 1751-8121.
\newblock \doi{10.1088/1751-8121/ab8b39}.

\bibitem[Hass et~al.(2023)Hass, {Carroll-Godfrey}, Corwin, and
  Corwin]{hassAnomalousFluctuationsExtremes2023b}
Jacob~B. Hass, Aileen~N. {Carroll-Godfrey}, Ivan Corwin, and Eric~I. Corwin.
\newblock Anomalous {{Fluctuations}} of {{Extremes}} in {{Many-Particle
  Diffusion}}.
\newblock \emph{Physical Review E}, 107\penalty0 (2):\penalty0 L022101,
  February 2023.
\newblock \doi{10.1103/PhysRevE.107.L022101}.

\bibitem[Hass et~al.(2024{\natexlab{a}})Hass, Corwin, and
  Corwin]{hassFirstpassageTimeManyparticle2024}
Jacob~B. Hass, Ivan Corwin, and Eric~I. Corwin.
\newblock First-passage time for many-particle diffusion in space-time random
  environments.
\newblock \emph{Physical Review E}, 109\penalty0 (5):\penalty0 054101, May
  2024{\natexlab{a}}.
\newblock \doi{10.1103/PhysRevE.109.054101}.

\bibitem[Das et~al.(2023{\natexlab{a}})Das, Drillick, and
  Parekh]{dasKPZEquationLimit2023b}
Sayan Das, Hindy Drillick, and Shalin Parekh.
\newblock {{KPZ}} equation limit of sticky {{Brownian}} motion.
\newblock August 2023{\natexlab{a}}.
\newblock \doi{10.48550/arXiv.2304.14279}.

\bibitem[See()]{SeeSupplementalMateriala}
See {{Supplemental Material}} at {{XXXX}}.

\bibitem[Parekh(2024)]{parekhInvariancePrincipleKPZ2024a}
Shalin Parekh.
\newblock Invariance principle for the {{KPZ}} equation arising in stochastic
  flows of kernels.
\newblock January 2024.
\newblock \doi{10.48550/arXiv.2401.06073}.

\bibitem[Schuss et~al.(2019)Schuss, Basnayake, and
  Holcman]{schussRedundancyPrincipleRole2019a}
Z.~Schuss, K.~Basnayake, and D.~Holcman.
\newblock Redundancy {{Principle}} and the {{Role}} of {{Extreme Statistics}}
  in {{Molecular}} and {{Cellular Biology}}.
\newblock \emph{Physics of Life Reviews}, 28:\penalty0 52--79, March 2019.
\newblock ISSN 1571-0645.
\newblock \doi{10.1016/j.plrev.2019.01.001}.

\bibitem[Linn and Lawley(2022)]{linnExtremeHittingProbabilities2022a}
Samantha Linn and Sean~D. Lawley.
\newblock Extreme {{Hitting Probabilities}} for {{Diffusion}}.
\newblock \emph{Journal of Physics A: Mathematical and Theoretical},
  55\penalty0 (34):\penalty0 345002, August 2022.
\newblock ISSN 1751-8121.
\newblock \doi{10.1088/1751-8121/ac8191}.

\bibitem[Lawley(2020{\natexlab{a}})]{lawleyDistributionExtremeFirst2020a}
Sean~D. Lawley.
\newblock Distribution of {{Extreme First Passage Times}} of {{Diffusion}}.
\newblock \emph{Journal of Mathematical Biology}, 80\penalty0 (7):\penalty0
  2301--2325, June 2020{\natexlab{a}}.
\newblock ISSN 0303-6812, 1432-1416.
\newblock \doi{10.1007/s00285-020-01496-9}.

\bibitem[Lawley(2020{\natexlab{b}})]{lawleyUniversalFormulaExtreme2020a}
Sean~D. Lawley.
\newblock Universal {{Formula}} for {{Extreme First Passage Statistics}} of
  {{Diffusion}}.
\newblock \emph{Physical Review E}, 101\penalty0 (1):\penalty0 012413, January
  2020{\natexlab{b}}.
\newblock \doi{10.1103/PhysRevE.101.012413}.

\bibitem[Madrid and Lawley(2020)]{madridCompetitionSlowFast2020a}
Jacob~B. Madrid and Sean~D. Lawley.
\newblock Competition between {{Slow}} and {{Fast Regimes}} for {{Extreme First
  Passage Times}} of {{Diffusion}}.
\newblock \emph{Journal of Physics A: Mathematical and Theoretical},
  53\penalty0 (33):\penalty0 335002, July 2020.
\newblock ISSN 1751-8121.
\newblock \doi{10.1088/1751-8121/ab96ed}.

\bibitem[Basnayake et~al.(2019)Basnayake, Schuss, and
  Holcman]{basnayakeAsymptoticFormulasExtreme2019a}
K.~Basnayake, Z.~Schuss, and D.~Holcman.
\newblock Asymptotic {{Formulas}} for {{Extreme Statistics}} of {{Escape
  Times}} in 1, 2 and 3-{{Dimensions}}.
\newblock \emph{Journal of Nonlinear Science}, 29\penalty0 (2):\penalty0
  461--499, April 2019.
\newblock ISSN 1432-1467.
\newblock \doi{10.1007/s00332-018-9493-7}.

\bibitem[Hass et~al.(2024{\natexlab{b}})Hass, Drillick, Corwin, and
  Corwin]{hassUniversalKPZFluctuations2024}
Jacob Hass, Hindy Drillick, Ivan Corwin, and Eric Corwin.
\newblock Universal {{KPZ Fluctuations}} for {{Moderate Deviations}} of
  {{Random Walks}} in {{Random Environments}}.
\newblock \emph{In preparation}, 2024{\natexlab{b}}.

\bibitem[Hass(2024)]{hassSuperUniversalBehaviorExtreme2024a}
Jacob Hass.
\newblock Super-{{Universal Behavior}} for {{Extreme Diffusion}} in {{Random
  Environments}}.
\newblock \emph{In preparation}, 2024.

\bibitem[Kardar et~al.(1986)Kardar, Parisi, and
  Zhang]{kardarDynamicScalingGrowing1986a}
Mehran Kardar, Giorgio Parisi, and Yi-Cheng Zhang.
\newblock Dynamic {{Scaling}} of {{Growing Interfaces}}.
\newblock \emph{Physical Review Letters}, 56\penalty0 (9):\penalty0 889--892,
  March 1986.
\newblock \doi{10.1103/PhysRevLett.56.889}.

\bibitem[Corwin(2012)]{corwinKardarParisiZhang2012a}
Ivan Corwin.
\newblock The {{Kardar}}--{{Parisi}}--{{Zhang Equation}} and {{Universality
  Class}}.
\newblock \emph{Random Matrices: Theory and Applications}, 01\penalty0
  (01):\penalty0 1130001, January 2012.
\newblock ISSN 2010-3263.
\newblock \doi{10.1142/S2010326311300014}.

\bibitem[Das et~al.(2023{\natexlab{b}})Das, Drillick, and
  Parekh]{dasKPZEquationLimit2023c}
Sayan Das, Hindy Drillick, and Shalin Parekh.
\newblock {{KPZ}} equation limit of random walks in random environments.
\newblock November 2023{\natexlab{b}}.
\newblock \doi{10.48550/arXiv.2311.09151}.

\bibitem[Amir et~al.(2011)Amir, Corwin, and
  Quastel]{amirProbabilityDistributionFree2011a}
Gideon Amir, Ivan Corwin, and Jeremy Quastel.
\newblock Probability {{Distribution}} of the {{Free Energy}} of the
  {{Continuum Directed Random Polymer}} in 1+1 {{Dimensions}}.
\newblock \emph{Communications on Pure and Applied Mathematics}, 64\penalty0
  (4):\penalty0 466--537, April 2011.
\newblock ISSN 00103640.
\newblock \doi{10.1002/cpa.20347}.

\bibitem[Calabrese et~al.(2010)Calabrese, Doussal, and
  Rosso]{calabreseFreeEnergyDistributionDirected2010a}
P.~Calabrese, P.~Le Doussal, and A.~Rosso.
\newblock Free-{{Energy Distribution}} of the {{Directed Polymer}} at {{High
  Temperature}}.
\newblock \emph{EPL (Europhysics Letters)}, 90\penalty0 (2):\penalty0 20002,
  April 2010.
\newblock ISSN 0295-5075.
\newblock \doi{10.1209/0295-5075/90/20002}.

\bibitem[Kardar(1987)]{kardarReplicaBetheAnsatz1987}
Mehran Kardar.
\newblock Replica {{Bethe}} ansatz studies of two-dimensional interfaces with
  quenched random impurities.
\newblock \emph{Nuclear Physics B}, 290:\penalty0 582--602, January 1987.
\newblock ISSN 0550-3213.
\newblock \doi{10.1016/0550-3213(87)90203-3}.

\bibitem[Bertini and Cancrini(1995)]{bertiniStochasticHeatEquation1995a}
Lorenzo Bertini and Nicoletta Cancrini.
\newblock The {{Stochastic Heat Equation}}: {{Feynman-Kac Formula}} and
  {{Intermittence}}.
\newblock \emph{Journal of Statistical Physics}, 78\penalty0 (5):\penalty0
  1377--1401, March 1995.
\newblock ISSN 1572-9613.
\newblock \doi{10.1007/BF02180136}.

\bibitem[Brockington and Warren(2022)]{brockingtonEdgeCloudBrownian2022a}
Dom Brockington and Jon Warren.
\newblock At the edge of a cloud of {{Brownian}} particles.
\newblock August 2022.
\newblock \doi{10.48550/arXiv.2208.11952}.

\bibitem[Norris(1997)]{norrisMarkovChains1997}
J.~R. Norris.
\newblock \emph{Markov {{Chains}}}.
\newblock Cambridge {{Series}} in {{Statistical}} and {{Probabilistic
  Mathematics}}. Cambridge University Press, Cambridge, 1997.
\newblock ISBN 978-0-521-63396-3.
\newblock \doi{10.1017/CBO9780511810633}.

\end{thebibliography}

\clearpage
\onecolumngrid
{\centering {\huge \textbf{Supplemental Material} \par}}

\onecolumngrid
\setcounter{equation}{0}
\renewcommand{\theequation}{S\arabic{equation}}

\section{Examples of Distributions $\randMeasure$}\label{sec:DistributionDefinitions}

The \emph{Dirichlet} distribution for $\nu$ is specified by a choice of $k\in \mathbb{Z}_{\geq 1}$ and $\vec{\alpha} = (\alpha_{-k},\ldots,\alpha_{k})\in \mathbb{R}_{>0}^{2k+1}$. It assigns probability density proportional to  $\prod_{j=-k}^{k} \xi(j)^{\alpha_j}$ to each vector $\xi=(\xi(-k),\ldots, \xi(k)) \in \mathbb{R}_{ \geq 0}^k$ satisfying $\sum_{j=-k}^{k} \xi(j) = 1.$ Then, we set $\xi(i) = 0$ for $i \notin [-k, k]$ so that $\xi$ is defined on $\Z$. Note that a sample $\xi$ under $\nu$ defines a probability distribution on $\mathbb{Z}$ due to the imposed normalization and non-negativity. When $\alpha_j\equiv 1$ for all $-k \leq j \leq k$, we call the resulting Dirichlet distribution \emph{uniform}$(k)$ since $(\xi(-k),\ldots, \xi(k))$ is uniformly distributed over vectors in $ \mathbb{R}_{ \geq 0}^k$ that sum up to $1$.

The \emph{normalized i.i.d.} distribution for $\nu$ is specified by a choice of $k\in \mathbb{Z}_{\geq 1}$ and a choice of probability distribution on $\mathbb{R}_{\geq 0}$. Let $X_{-k},\ldots, X_{k}$ be chosen independently according to the specified probability distribution, and then define $\xi(i) = X_i (\sum_{j=-k}^{k} X_j)^{-1}$ for $i \in [-k, k]$ and $\xi(i) = 0$ otherwise. Such $\xi$ are normalized to sum to 1 and are non-negative, and hence define a probability distribution on $\mathbb{Z}$. When the $X_i$ are Gamma distributed with parameter $\alpha$, the resulting measure on $\xi$ matches the Dirichlet distribution with $\alpha_i\equiv\alpha$.


The \emph{random delta} distribution for $\nu$ is specified by $k \in \Z_{\geq 1}$. Two numbers $X_1,X_2$ are drawn uniformly without replacement from $\{-k,\ldots, k\}$. We set $\xi(X_1)=\xi(X_2)=1/2$ and $\xi(i)=0$ for all $i \neq X_1, X_2$.

\section{Numerical Methods}

We describe how we numerically measure the mean and variance of $\envFPT$, $\samFPT$ and $\min$. We begin by numerically computing the probability mass function of the first passage time for a single particle (we drop the superscript and just call it $R(t)$), defined as $\tauL = \mathrm{min}(t : R(t) \geq L)$. To do so, we consider $R_L(t) = R(\mathrm{min}(t, \tauL))$, the random walk stopped (or absorbed) when $R(t) \geq L$. We denote the probability mass function of $R_L(t)$ as $p^{\env}_L(x,t) = \mathbb{P}^{\env}(R_L(t) = x)$. Given an environment $\env$, $p^{\env}_L(x,t)$ uniquely solves
\begin{equation}\label{eq:FPTRecursionRelation}
    p^{\env}_L(x,t+1) = \sum_{i < L} p^{\env}_L(i, t) \xi_{t, i}(x-i)
\end{equation}
for $x \in (-\infty, L) \cap \Z$ and $t\geq 0$, subject to the absorbing boundary condition 
\begin{equation}\label{eq:AbsorbingBoundary}
    p^{\env}_L(L,t+1) = p^{\env}_L(L, t) + \sum_{i < L} p^{\env}_L(i, t) \sum_{j \geq L} \xi_{t, i}(j-i)
\end{equation}
and initial condition $p^{\env}_L(x,t) = \mathbbm{1}_{x=0}$.
The probability of $\tauL$ occurring before time $t$ is given by the probability of $R_L(t)$ being absorbed before time $t$, which is to say
\begin{equation}\label{eq:FPTDist}
    \mathbb{P}^{\env}(\tauL \leq t) = p^{\env}_L(L,t).
\end{equation}

For a given environment distribution $\nu$, we numerically sample $\randMeasure$ to generate an environment, $\env$, and then we use this environment to compute $\mathbb{P}^{\env}(\tauL \leq t)$ via \eqref{eq:FPTRecursionRelation} and \eqref{eq:AbsorbingBoundary}. We measure $\envFPT$ in this environment by finding the minimum time $t$ such that $\mathbb{P}^{\env}(\tauL \leq t) \geq 1/N$. We then numerically compute the distributions of $\min$ and $\samFPT$ in the given environment. For $\min$ we use \eqref{eq:MinDef} and for $\samFPT$ we combine \eqref{eq:MinDef} and our definition $\samFPT := \min - \envFPT$ and thus compute the distribution of $\samFPT$ as 
\begin{equation}\label{eq:samDist}
    \mathbb{P}^{\env}(\samFPT \leq t) = 1 - \left(1 - \mathbb{P}^{\env}(\tauL \leq t + \envFPT) \right)^N .
\end{equation}
Since $N$ is quite large, we use arbitrary precision floating point arithmetic to compute the $N^{th}$ power in \eqref{eq:MinDef} and  \eqref{eq:samDist}. Now, given these computed distribution functions we compute $\expEnv[\min]$ and $\varEnv(\min)$, $\expEnv[\samFPT]$ and $\varEnv(\samFPT)$. Finally, we repeat this procedure for many different samples of $\env$. We compute  $\expTotal{\envFPT}$ and $\varTotal{\envFPT}$ by taking the mean and variance of $\envFPT$ over these samples. For $\min$ and $\samFPT$ we utilize the law of total expectation, i.e., $\expTotal{\min} = \expAnnealed{\expEnv[\min]}$ (likewise for $\samFPT$) and the total law of variance 
$\varTotal{\min} = \varAnnealed\left(\expEnv\left[\min\right]\right) + \expAnnealed{\varEnv\left(\min\right)}$.

We use a nearly identical procedure to numerically compute the mean and variance of the corresponding extreme location quantities $\envMax$, $\samMax$ and $\max$. Given an environment $\env$ we start by numerically computing $p^{\env}(x,t)$ using the recurrence relation given in \eqref{eq:RecussionRelation}. We then calculate $\envMax$ by finding the maximum position $x$ such that $\mathbb{P}^{\env}(R(t) \geq x) \geq 1/N$. We compute the distribution of $\max$ using \eqref{eq:MaxDef}, and the distribution of $\samMax$ by combining \eqref{eq:MaxDef} and our definition $\samMax := \max - \envMax$ to find
\begin{equation}\label{eq:samMaxDef}
    \mathbb{P}^{\env}(\samMax \leq x) = (1 - \mathbb{P}^{\env}(R(t) \leq x + \envMax))^N.
\end{equation}
From these we compute $\expEnv[\max]$, $\varEnv(\max)$, $\expEnv[\samMax]$ and $\varEnv(\samMax)$. Finally, by repeating for several samples of $\env$, as above, we compute $\expTotal{\envMax}$ and $\varTotal{\envMax}$, and then
$\mathbf{E}[\max]$, $\mathbf{E}[\samMax]$, $\varTotal{\max}$ and $\varTotal{\samMax}$.

We numerically measure $\envFPT$, $\samFPT$ and $\min$ for several distributions. We simulate a Dirichlet distribution with $\vec{\alpha} = (12, 1, 12)$ which is peaked at $-1$ and $1$ with particles having a small probability of staying at the same location. We also simulate a Dirichlet distribution with $\vec{\alpha} = (2, 1, 1/4, 4, 1/2)$ to study a distribution that is not symmetric in the average environment. We consider a uniform distribution on the interval $[-k, k]$ for $k = 1, 2, 5$. Recall that the uniform distribution is a special case of the Dirichlet distribution with all $\alpha_i = 1$. Lastly, we consider the random delta distribution on the interval $[-k, k]$ for $k = 1, 2, 5$.


\section{Convergence to the Stochastic Heat Equation}
We begin by summarizing the results in the forthcoming work \cite{hassUniversalKPZFluctuations2024}. These results also follow from \cite{parekhInvariancePrincipleKPZ2024a}. We show that for $N \in \Z_{> 0}$, $T \in N^{-1} \Z_{\geq 0}$ and $X \in (2DN)^{-1/2} (\Z - c_N T)$, the scaled moderate deviations of the tail probability for a single random walker converges as $N\rightarrow \infty$:
\begin{equation}\label{eq:tailResult}
    \frac{N^{1/4} C_{N, T, X}}{\sqrt{2D}} \mathbb{P}^{\env}( R^1(NT) \geq \centeringTerm T  + \sigma N^{1/2} X) \Rightarrow \tilde{Z}(X, T)
\end{equation}
with scaling parameters
\begin{equation}
    C_{N, T, X} = \frac{\mathrm{exp}\left\{\displaystyle\frac{\centeringTerm}{2D N^{1/4}} T + \frac{1}{\sqrt{2D}} N ^{1/4} X\right\}}{\left(\displaystyle\sum_{i\in\Z} \meanOmega{i} \displaystyle \mathrm{exp}\left\{ \frac{i}{2DN^{1/4}}\right\} \right)^{NT}}\,\,, \qquad \qquad\centeringTerm = N^{3/4} + \frac{\sum_{i\in Z} \meanOmega{i} i^3}{2 (2D)^2} N^{1/2}.
\end{equation}
The convergence is shown in \cite{hassUniversalKPZFluctuations2024} at the level of the first two moments (stronger process-level convergence is shown in \cite{parekhInvariancePrincipleKPZ2024a}), and the limiting process $\tilde{Z}(X, T)$ is the solution to the multiplicative Stochastic Heat Equation (mSHE) with $\tilde{Z}(X,0)=\delta(X)$ initial data: 
\begin{equation}
    \partial_T \tilde{Z} = \frac{1}{2} \partial_X^2 \tilde{Z} + \sqrt{\frac{2\extCoef}{(2 D)^{3/2}}} \tilde{Z} \eta. 
\end{equation}
Here $\eta(X,T)$ is a space-time Gaussian white noise (i.e. $\mathbb{E}[\eta(X,T)] = 0$ and $\mathbb{E}[\eta(X,T) \eta(X', T')] = \delta(X-X')\delta(T-T')$ where $\delta$ is the Dirac delta function). The noise strength is controlled by the Einstein diffusion coefficient $D$ defined in \eqref{eq:edc} and 
$\extCoef$. 

At the full level of generality of RWRE models we introduced in the \emph{RWRE models} section, \cite{hassUniversalKPZFluctuations2024} (see also \cite{parekhInvariancePrincipleKPZ2024a}) provides a somewhat involved formula for $\extCoef$:
\begin{equation}\label{eq:lambdaFullDef}
    \extCoef := \frac{\driftVar}{\sum_{l=0}^{\infty} \absMeasure{l}\expTotal{\diffRandomWalk{t+1} - \diffRandomWalk{t} \mid \diffRandomWalk{t} = l}}.
\end{equation}
The numerator here is defined in terms of $Y$, a random jump distributed according to $\xi$. As 
in \eqref{eq:dExt}, this numerator is equal to $2\dExt$. The denominator requires more explanation. Consider two random walks $R^1$ and $R^2$ distributed according to the measure $\mathbf{P}$ (i.e., after integrating over the law $\nu$ of the environment). These walks are not independent as they are coupled to the same (integrated out) environment---we call these the \emph{two-point motions} as they have the law of two tracer particles when the environment is hidden. Define the gap between them to be 
$$
\Delta(t) := |R^1(t)-R^2(t)|.
$$
There exists a unique (infinite mass) invariant measure for $V(t):=R^1(t)-R^2(t)$ and let $\mu(l)$ be the mass assigned to $l\in \Z$ with the normalization that $\mu(0)=1$ (see e.g., \cite{norrisMarkovChains1997} for background on invariant measures for Markov chains). The corresponding invariant measure for $\Delta(t)$ is therefore 
\begin{equation}\label{eq:absMeasureDef}
\absMeasure{l} :=  \begin{cases} 
        1 & \text{if $l=0$} \\
       2 \mu(l) &  \text{if $l >0$} 
    \end{cases}.
\end{equation}
This invariant measure can be understood physically as follows. Start two particles near each other and let them diffuse in their common environment. After a relatively long time, measure the distance between them. Repeat this for many different environments, thus building up a histogram of the distances between these two-point motions. The typical distance will be large, but if we cutoff to consider relatively short distances, and normalize the histogram to put weight $1$ at distance $0$, then it will converge to $\absMeasure{l}$. A slightly different way that this same measure should arise is from surveying the inter-particle distances over all particles in a many-particle diffusion. Normalizing the histogram of these distances to be $1$ at distance $0$, will yield a histogram that converges to $\absMeasure{l}$ at distance $l$. This should hold for a single environment since the inter-particle distances (on the short-scale that we are considering) will generally feel different parts of the environment, and hence experience some averaging.

The denominator in \eqref{eq:lambdaFullDef} is independent of $t$ and it has the interpretation as the expected change in the two-point motion gap when started under its invariant measure. For large enough $l$ (since we have assumed a finite range for the jump distribution) the expected change of the gap will be zero and hence the sum will be a finite one. 

We will show that $\extCoef$ simplifies considerably for a wide class of distributions $\nu$ in Section \ref{sec:computingCoeffs}, and (before that) in the next section we will explain (in the spirit of \cite{hassAnomalousFluctuationsExtremes2023b,hassFirstpassageTimeManyparticle2024}) how we go from this mSHE convergence result to our extreme diffusion theoretical predictions.

However, let us first briefly summarize the approach used in \cite{hassUniversalKPZFluctuations2024} to derive this mSHE convergence result. First of all, \eqref{eq:tailResult} is only demonstrated therein at the level of convergence of first and second moments. Convergence of higher moments follows similarly as noted below. Since the moments of the mSHE do not characterize its distribution, more work is needed, as in \cite{parekhInvariancePrincipleKPZ2024a}, to show convergence in distribution, or convergence of the space-time process. 

The second moment of the LHS in \eqref{eq:tailResult} can be expressed via a discrete Feynman-Kac formula in terms of the expectation of the exponential of the self-intersection time for the two-point motion $(R^1,R^2)$ under a tilting of the measure $\mathbf{P}$ (high integer moments involve higher order self-intersection times). For the mSHE, all integer moments can be expressed in terms of the expectation of the exponential of the pair local times at zero for independent Brownian motions (this is the replica method, see \cite{kardarReplicaBetheAnsatz1987, bertiniStochasticHeatEquation1995a}). The $k$-point motion converges diffusively to $k$ independent Brownian motions, yet the discrete self-intersection time does not converge to the local time at zero for the Brownian motions. 

This failure of self-intersection time to local time convergence may seem surprising, but a very simple example should convince the reader that such convergence is more subtle. For instance, imagine two independent SSRWs, one started at 0 and one started at 1. They converge jointly to two Brownian motions, yet their intersection time is always 0 (they are on different sublattices) and hence does not converge to the local time at 0 for the Brownian motions. 

A discrete version of the Tanaka formula---which, for a  Brownian motion $B$, shows that $|B(t)|= \int_0^t \mathrm{sgn}(B(s)) dB(s) + L(t)$ where $\mathrm{sgn}(x)$ is the derivative of the absolute value function $|x|$ (set to be 0 at 0) and where $L(t)$ is the local time at zero up to time $t$---is used in \cite{hassUniversalKPZFluctuations2024} to identify the discrete quantity that converges to the Brownian local time $L(t)$. That discrete local time involves reweighting the expected change in distance between the two-point motion according to its invariant measure. This allows us to identify the limit of the self-intersection time and leads to the denominator in \eqref{eq:lambdaFullDef}.

\section{Approximating the Tail Probability}
After taking logarithms in \eqref{eq:tailResult}, we see that as $N \rightarrow \infty$
\begin{equation}\label{eq:logTail}
    \ln(\mathbb{P}^{\env}( R^1(NT) \geq \centeringTerm T)) \approx - \frac{c_N T}{2 D N^{1/4}}  - \frac{N^{1/4}X}{\sqrt{2D}} + NT \ln\left(\displaystyle\sum_{i\in\Z} \meanOmega{i} \displaystyle \mathrm{exp}\left\{ \frac{i}{2DN^{1/4}}\right\}\right) - \ln\left( \frac{N^{1/4}}{\sqrt{2D}}\right) + \tilde{h}(X, T)
\end{equation}
where $\tilde{h}(X,T) = \ln(\tilde{Z}(X,T))$ solves the KPZ equation with narrow wedge initial data and with $\extCoef$-dependent noise strength (we now also include tildes on the $X$ and $T$ variables to simplify the below change of variables), 
$$
    \partial_{\tilde T} \tilde{h} = \frac{1}{2} \partial_{\tilde X}^2 \tilde{h} + \frac{1}{2} (\partial_{\tilde X} \tilde{h})^2 + \sqrt{\frac{2\extCoef}{(2 D)^{3/2}}} \eta .
$$
Defining $h(X,T)=\tilde{h}(\tilde X,\tilde T)$ with $T = \frac{4 \extCoef^2}{(2D)^{3}} \tilde T$ and $X = \frac{2 \extCoef}{(2D)^{3/2}} \tilde X$, $h$ solves the standard coefficient KPZ equation
$$
    \partial_{T}h= \frac{1}{2} \partial_{X}^2 h + \frac{1}{2} (\partial_{X} h)^2 + \eta.
$$
We also have that as $N \rightarrow \infty$,
\begin{equation*}
    \left(\displaystyle\sum_{i\in\Z} \meanOmega{i} \displaystyle \mathrm{exp}\left\{ \frac{i}{2DN^{1/4}}\right\}\right) \approx \frac{N^{-1/2}}{4 D} + \frac{\sum_{i\in\Z}\meanOmega{i}i^3 }{6 (2 D)^3} N^{-3/4} + \mathcal{O}\left( N^{-1} \right) . 
\end{equation*}
Substituting this into \eqref{eq:logTail} and using our transformation to the KPZ equation, we find 
\begin{align*}
    \ln\left(\mathbb{P}^{\env}\left(R^1(NT) \geq N^{3/4} T + \frac{m_3 N^{1/2}T}{2 (2D)^2} + \sqrt{2DN} X\right)\right) \\ 
    \approx -\frac{N^{1/2} T}{4 D} - \frac{N^{1/4}X}{\sqrt{2D}} &- \frac{N^{1/4} T m_3}{3 (2D)^3} + \log\left(\frac{N^{1/4}}{\sqrt{2D}} \right) + h\left(\frac{4 \extCoef^2}{(2D)^{3}} T, X\right) + \mathcal{O}(T)
\end{align*}
where $m_3 = \sum_{i \in \Z} \meanOmega{i} i^3$. We now introduce the time $t \coloneqq NT$, velocity $v \coloneqq T^{1/4}$, and rescaled position $y= \frac{X}{v^2}$. Making these substitutions, we find 
\begin{align}\label{eq:tailTransform}
\begin{split}
    \ln\left( \mathbb{P}^{\env}\left(R^1(t) \geq vt^{3/4} + \frac{m_3}{2(2D)^2}v^2 t^{1/2} + \sqrt{2Dt} y \right) \right) \\
    \approx - \frac{v^2}{4 D} t^{1/2} -\frac{v y}{\sqrt{2D}} t^{1/4}  - \frac{v^3 m_3}{3 (2D)^3} t^{1/4}& + \log\left(\frac{t^{1/4}}{\sqrt{2D} v}\right) + h\left(\frac{4 \extCoef^2}{(2D)^{3}} v^4, yv^2\right) + \mathcal{O}(v^4).
\end{split}
\end{align}

\section{Extreme First Passage Time Theoretical Predictions}

Here we derive asymptotic predictions for the means and variances of $\envFPT$, $\samFPT$ and $\min$. Much of this analysis follows from \cite{hassFirstpassageTimeManyparticle2024}, where they derived predictions for the nearest neighbor case with a uniform distribution.  

We start by substituting $y = 0$ and $L = v t^{3/4}$ into \eqref{eq:tailTransform} such that 
\begin{equation}
    \ln\left( \mathbb{P}^{\env}\left(R^1(t) \geq L + \frac{m_3 L^2}{2(2D)^2 t} \right) \right) \approx - \frac{L^2}{4 D t}  - \frac{m_3 L^3}{3 (2D)^3 t^2} + \log\left(\frac{t}{\sqrt{2D} L}\right) + h\left(\frac{4 \extCoef^2 L^4}{(2D)^{3} t^3}, 0 \right) + \mathcal{O}\left(\frac{L^4}{t^3}\right).
\end{equation}
We can now drop all subdominant terms as they will not contribute to the asymptotic predictions (though they could offer some higher-order corrections that we do not probe here). This yields a rougher approximation where we do not track the order of the error
\begin{equation}\label{eq:tailProbL}
    \ln\left( \mathbb{P}^{\env}\left(R^1(t) \geq L\right) \right) \approx - \frac{L^2}{4 D t}  + h\left(\frac{4 \extCoef^2 L^4}{(2D)^{3} t^3}, 0 \right).
\end{equation}
We now utilize the non-backtracking approximation $\mathbb{P}^{\env}(R^1(t) \geq L) \approx \mathbb{P}^{\env}(\tauL \leq t)$ (as discussed in \cite{hassFirstpassageTimeManyparticle2024}) as $L$ gets large to yield
\begin{equation}\label{eq:tailTaul}
    \ln\left( \mathbb{P}^{\env}\left(\tauL \leq t \right) \right) \approx - \frac{L^2}{4 D t}   + h\left(\frac{4 \extCoef^2 L^4}{(2D)^{3} t^3}, 0 \right).
\end{equation}
We now substitute $\envFPT$ into this equation, recalling that $\envFPT$ is approximately the time $t$ such that $\mathbb{P}^{\env}(\tauL \leq t) = 1 / N$ (in fact, the minimum time $t$ satisfying $\mathbb{P}^{\env}(\tau_L \leq t) \geq 1/N$ though this difference is negligible):
\begin{equation}\label{eq:fptLnN}
    -\ln\left(N \right) \approx - \frac{L^2}{4 D \envFPT}  + h\left(\frac{4 \extCoef^2 L^4}{(2D)^{3} \left(\envFPT\right)^3}, 0 \right).
\end{equation}
We can solve this equation perturbatively for $\envFPT$. The term $-\frac{L^2}{4D \envFPT}$ in \eqref{eq:fptLnN} is dominant for large $L$, which yields the first-order estimate 
\begin{equation}\label{eq:T0}
    \envFPT \approx T^N_L \coloneqq \frac{L^2}{4D \ln(N)}.
\end{equation}
We now consider a small perturbation about $T^N_L$ such that $\envFPT = T^N_L + \delta$ where $\delta \ll T^N_L$ contains the randomness of $\envFPT$. Substituting this into \eqref{eq:fptLnN}, we find 
\begin{equation}\label{eq:envFPTApprox}
    -\ln\left(N \right) \approx - \frac{L^2}{4 D (T^N_L + \delta)}  + h\left(\frac{4 \extCoef^2 L^4}{(2D)^{3} \left(T^N_L + \delta\right)^3}, 0 \right).
\end{equation}
Since $\delta \ll T^N_L$, we approximate $-\frac{L^2}{4D(T^N_L + \delta)} \approx -\frac{L^2}{4 D T^N_L} + \frac{L^2}{4D(T^N_L)^2}\delta$ and $\frac{4 \extCoef^2 L^4}{(2D)^{3} \left(T^N_L + \delta\right)^3}\approx \frac{4 \extCoef^2 L^4}{(2D)^{3} \left(T^N_L\right)^3}$. As such we can solve for $\delta$, yielding 
\begin{align*}
    \delta &\approx - \frac{L^2}{4 D \ln(N)^2} \,\cdot\,h\left(\frac{4 \extCoef^2 L^4}{(2D)^{3} (T^N_L)^3}, 0 \right)=  -\frac{L^2}{4D \ln(N)^2} \,\cdot\, h\left( \frac{32 \extCoef^2 (\ln(N))^3}{L^2}, 0\right).
\end{align*}
Therefore, the mean and variance of $\envFPT$ is given by 
\begin{align}
    \expTotal{\envFPT} &\approx \frac{L^2}{4D \ln(N)} \\ 
    \varTotal{\envFPT} &\approx \frac{L^4}{(4D)^2 \ln(N)^4} \varTotal{h\left(\frac{32 \extCoef^2 (\ln(N))^3}{L^2}, 0 \right)} \label{eq:envFPTKPZ}.
\end{align}
We now consider the limit where $L \gg (\ln(N))^{3/2}$. 
In this limit, we simplify \eqref{eq:envFPTKPZ} using the small-time KPZ Gaussian approximation 
\begin{equation}\label{eq:KPZapprox}
    h(s, 0) \approx -\frac{s}{24} - \ln\left( \sqrt{2 \pi s} \right) + \left( \frac{\pi s}{4} \right)^{1/4} G_s
\end{equation}
where $G_s$ converges as $s\rightarrow 0$ to a standard Gaussian, see for instance \cite{amirProbabilityDistributionFree2011a,calabreseFreeEnergyDistributionDirected2010a}. Thus, for $L \gg \extCoef(\ln(N))^{3/2}$ we find
\begin{equation}
    \varTotal{\envFPT} \approx \extCoef \frac{\sqrt{2\pi} L^3}{8 D^2 \log(N)^{5/2}}.
\end{equation}


We now derive the distribution, and subsequently the mean and variance, for the randomness due to sampling random walks, $\samFPT$. We begin by using the approximation, $(1 + x)^N \approx e^{xN}$ for $x \ll 1$ and $N \rightarrow \infty$. Therefore, for $N \rightarrow \infty$ and small $\mathbb{P}^{\env}(\tau_L \leq t)$, we approximate \eqref{eq:samDist} as 
\begin{equation}
    \ln\left(\mathbb{P}^{\env}(\samFPT > t)\right) \approx - N \mathbb{P}^{\env}(\tau_L \leq t +\envFPT).
\end{equation}
Substituting \eqref{eq:tailTaul} yields
\begin{equation}
    \ln\left(\mathbb{P}^{\env}(\samFPT > t)\right) \approx -N \mathrm{exp}\left\{-\frac{L^2}{4D(t + \envFPT)}\right\}
\end{equation}
where we have dropped all but the leading order term. We now assume $\envFPT \gg t$, which is justified because, as we will show, $\expTotal{\envFPT} \gg \expTotal{\samFPT}$. When $\envFPT \gg t$, we approximate $(t + \envFPT)^{-1} \approx \frac{1}{\envFPT}\left(1 - \frac{t}{\envFPT}\right)$ such that
\begin{equation}\label{eq:smallSamFPTApprox}
    \ln\left(\mathbb{P}^{\env}(\samFPT > t)\right) \approx  - N \mathrm{exp} \left\{ -\frac{L^2}{4D\envFPT} -\frac{t L^2}{4D \left(\envFPT\right)^2} \right\}.
\end{equation}
Now replacing $\envFPT$ with its leading order approximation, $T^N_L$, from  \eqref{eq:T0}, we find 
\begin{align*}
    \ln\left(\mathbb{P}^{\env}(\samFPT \leq t)\right) &\approx -\mathrm{exp} \left\{ -\frac{4D \ln(N)^2 t}{L^2} \right\}.
\end{align*}
By replacing $\envFPT$ with $T^N_L$ we have assumed that $\samFPT$ and $\envFPT$ are independent since $\samFPT$ no longer depends on the randomness of $\envFPT$. Though likely theoretically justifiable, this approximation is a fortiori justified by our numerics as shown in Figure \ref{fig:FPTMax}. From this we find that $-\samFPT$ is Gumbel distributed with mean and variance
\begin{align}
    \mathbf{E}[\samFPT] &\approx -\frac{\gamma L^2}{4 D \ln(N)^2} \\
    \mathbf{Var}(\samFPT) &\approx \frac{\pi^2 L^4}{96 D^2 \ln(N)^4}
\end{align}
where $\gamma \approx 0.577$ is the Euler-Mascheroni constant. Notice that as $L$ and $N$ tend to infinity with $L \gg \ln(N)^{3/2}$, $\expTotal{\samFPT} \ll \expTotal{\envFPT}$ which justifies the approximation used in \eqref{eq:smallSamFPTApprox}.

We solve for the mean and variance of $\min$ by rearranging our definition of $\samFPT$ such that $\min = \samFPT + \envFPT$. Since $\expTotal{\samFPT} \ll \expTotal{\envFPT}$, 
\begin{equation}
    \expTotal{\min} \approx \frac{L^2}{4D \ln(N)}.
\end{equation}
Since $\envFPT$ and $\samFPT$ are independent, $\varTotal{\min} = \varTotal{\samFPT} + \varTotal{\envFPT}$ such that 
\begin{equation}
    \varTotal{\min} \approx \frac{\pi^2 L^4}{96 D^2 \ln(N)^4} + \extCoef \frac{\sqrt{2\pi} L^3}{8 D^2 \log(N)^{5/2}}
\end{equation}
in the limit $L \gg \ln(N)^{3/2}$.

\section{Extreme Location Theoretical Predictions}



Here we derive asymptotic predictions for the means and variances of $\envMax$, $\samMax$ and $\max$. As above, much of this analysis follows from \cite{hassAnomalousFluctuationsExtremes2023b}, where they derived predictions for the nearest neighbor case with a uniform distribution. The argument proceeds similarly to the first passage time analysis. Starting at \eqref{eq:tailProbL}, we substitute $\envMax$ for the position $L$ such that  $\mathbb{P}^{\env}(R^1(t) \geq L) = 1/N$, thus yielding
\begin{equation}\label{eq:maxPartProb}
    -\ln\left( N \right) \approx - \frac{\left( \envMax \right)^2}{4 D t}   + h\left(\frac{4 \extCoef^2 \left( \envMax \right)^4}{(2D)^{3} t^3}, 0 \right).
\end{equation}
The first term on the right-hand side dominates and yields the first-order estimate 
\begin{equation}\label{eq:x0}
    \envMax \approx X^N_t \coloneqq \sqrt{4 D t \ln(N)}. 
\end{equation}
We consider a small perturbation, $\delta$, about $X^N_t$ such that $\envMax = X^N_t + \delta$ where $\delta \ll X^N_t$. Substituting this into \eqref{eq:maxPartProb} and only taking the highest order terms yields
\begin{equation}
    \delta \approx \sqrt{\frac{D t}{\ln(N)}} h \left( \frac{8 \extCoef^2 \ln(N)^2}{D t}, 0 \right).
\end{equation}
Therefore, the mean and variance of $\envMax$ satisfy
\begin{align}
    \expTotal{ \envMax } &\approx \sqrt{4D t \ln(N)} \\
    \varTotal{ \envMax } &\approx \frac{D t}{\ln(N)} \varTotal{ h \left( \frac{8 \extCoef^2 \ln(N)^2}{D t}, 0 \right) }. 
\end{align}
When $t \gg \frac{\extCoef^2}{D} \ln(N)^2$, we use the small argument expansion of the KPZ equation in  \eqref{eq:KPZapprox} to approximate 
\begin{equation}
    \varTotal{ \envMax } \approx \extCoef \sqrt{2\pi D t}.
\end{equation}

We now derive the distribution for the randomness due to sampling random walks, 
$\samMax$.
For $N\rightarrow \infty$ and small $\mathbb{P}^{\env}(R(t) \leq x + \envMax)$, we approximate \eqref{eq:samMaxDef} as
\begin{equation}
    \ln(\mathbb{P}^{\env}(\samMax \leq x)) \approx -N \mathbb{P}^{\env}(R(t) \leq x +\envMax).
\end{equation}
Substituting \eqref{eq:tailProbL} yields 
\begin{equation}
    \ln(\mathbb{P}^{\env}(\samMax \leq x)) \approx  -N \mathrm{exp}\left\{ -\frac{(x+\envMax)^2}{ 4 D t} \right\}
\end{equation}
where we have only kept the leading order term of $\mathbb{P}^{\env}(R(t) \leq x)$. We assume $\envMax \gg x$ 
which is justified because, as we will show, $\expTotal{\envMax} \gg \expTotal{\samMax}$. We use this to approximate $(x + \envMax)^2 \approx (\envMax)^2 + 2 \envMax x$ such that 
\begin{equation}\label{eq:smallSamApprox}
    \ln(\mathbb{P}^{\env}(\samMax \leq x)) \approx \mathrm{exp}\left\{ -N \mathrm{exp}\left\{ -\frac{(\envMax)^2}{ 4 D t} -\frac{\envMax x}{2 D t} \right\}\right\}.
\end{equation}
Replacing $\envMax$ with its first-order approximation, $X^N_t$, in \eqref{eq:x0}, we find 
\begin{equation}
    \ln(\mathbb{P}^{\env}(\samMax \leq x)) \approx  - \exp \left\{ -\sqrt{\frac{\ln(N)}{Dt}} x \right\}.
\end{equation}
Therefore, $\samMax$ is Gumbel distributed with mean and variance 
\begin{align}
    \mathbf{E}[\samFPT] \approx \gamma \sqrt{\frac{D t}{\ln(N)}} \\ 
    \mathbf{Var}(\samFPT) \approx \frac{\pi^2 D t}{6 \ln(N)}.
\end{align}
Notice that as $t \rightarrow \infty$, $\expTotal{\envMax} \gg \expTotal{\samMax}$ which justifies our approximation in \eqref{eq:smallSamApprox}.

We solve for the mean and variance of $\max$ by rearranging our definition of $\samMax$ such that $\max = \samMax + \envMax$. Since $\expTotal{\envMax} \gg \expTotal{\samMax}$,
\begin{equation}
    \expTotal{\max} \approx \sqrt{4 D t \ln(N)}.
\end{equation}
Since $\envMax$ and $\samMax$ are assumed to be asymptotically independent, $\varTotal{\max} = \varTotal{\envMax} + \varTotal{\samMax}$ so
\begin{equation}
    \varTotal{\max} \approx \frac{\pi^2 D t}{6 \ln(N)} + \sqrt{2\pi D t}
\end{equation}
when $t \gg \extCoef^2 \ln(N)^2$.

\section{Calculating the Coefficient for Several Distributions} \label{sec:computingCoeffs}
We demonstrate that the coefficient $\extCoef$ in the mSHE/KPZ equation limit  \eqref{eq:tailResult} simplifies to the expression in \eqref{eq:lambdaSimplified} for a class of distributions $\nu$ including those introduced earlier (i.e., the Dirichlet, normalized i.i.d., and random delta distributions), as well as all nearest neighbor distributions.

\subsection{General Model}
We study the following class of distributions such that:  
\begin{equation}\label{eq:modelCondition}
\textrm{There exists some } c \in (0, 1) \textrm{ such that for all } i\neq j\qquad 
    \expAnnealed{\xi(i)\xi(j)} = c \expAnnealed{\xi(i)} \expAnnealed{\xi(j)}.
\end{equation}  
We will also initially assume that the difference walk $V(t) = R^1(t) - R^2(t)$ (where $R^1$ and $R^2$ are distributed according to $\mathbf{P}$) is irreducible (i.e., $V(t)$ can reach any location on $\mathbb{Z}$ when started from $0$), although we will later also consider the nearest neighbor model in Section \ref{sec:NearestNeighbor}, which does not satisfy this condition as the difference walk $V(t)$ in that case is restricted to the even integer sublattice. A sufficient condition for $V(t)$ to be irreducible is that 
\begin{equation}\label{eq:irreducibility}
    \meanOmega{i} > 0 \textrm{ for all } i \in \{-1, 0, 1\}.
\end{equation}

We compute $\extCoef$ by simplifying the numerator and denominator of \eqref{eq:lambdaFullDef} in terms of $c$ and then matching this to \eqref{eq:lambdaSimplified}. The numerator of \eqref{eq:lambdaFullDef} simplifies to  
\begin{equation}\label{eq:generalDriftVar}
    \driftVar = \sum_{i\in\Z} (1-c) \expAnnealed{\xi(i)} i^2.
\end{equation}

Simplifying the denominator of \eqref{eq:lambdaFullDef} is much more involved. We do this in three steps. First, we compute the invariant measure, $\absMeasure{l}$  and show that  
\begin{equation}
    \absMeasure{l} = 
    \begin{cases}
        1
 &\text{if $i=0$} \\
  2c &\text{if $i \neq 0$}  \end{cases}.
\end{equation}
Then we simplify the expression  $\expTotal{\diffRandomWalk{t+1} - \diffRandomWalk{t} \mid \diffRandomWalk{t} = l}$ and show that (without using the assumption \eqref{eq:modelCondition})
\begin{equation}\label{eq:expSimplified}
    \expTotal{\diffRandomWalk{t+1} - \diffRandomWalk{t} \mid \diffRandomWalk{t} = l} = \begin{dcases} 
      \sum_{i, j \in \Z}|i-j|\expAnnealed{\xi(i) \xi(j)} & l=0 \\
      \sum_{|i-j| > l} (|i-j| - l) \meanOmega{i}\meanOmega{j} & l>0 \\
   \end{dcases}.
\end{equation}
Lastly, we combine these two results to compute the sum over $l$ and show that 
\begin{equation}
    \sum_{l=0}^{\infty}\expTotal{\diffRandomWalk{t+1} - \diffRandomWalk{t} \mid \diffRandomWalk{t} = l} \absMeasure{l} = 2c \sum_{i\in\Z} i^2 \meanOmega{i}.
\end{equation}

\emph{The invariant measure:} We now compute the invariant measure for the walk $V(t)$. The transition probabilities of $V$ are given by 
\begin{equation}\label{eq:jump_rates}
    p(i,j) := \mathbf{P}(V(t+1) = j \mid V(t) = i) = \begin{cases}
        \sum_{k \in \Z}\expAnnealed{\xi(k) \xi(k-j)}
 &\text{if $i=0$} \\
  \sum_{k \in \Z}\meanOmega{k}\meanOmega{k -j + i}   &\text{if $i \neq 0$}.  \end{cases}.
\end{equation}
Notice that if $i, j \neq 0$, then by a change of variables, we have 
\begin{align}
    p(i, j)&= \sum_{k \in \Z}\meanOmega{k -j + i} \meanOmega{k} \nonumber \\
    &= \sum_{\tilde{k} \in \Z} \meanOmega{\tilde{k}} \meanOmega{\tilde{k} - i + j } \nonumber \\
    &= p(j,i). \label{eq:pijEquality}
\end{align}
It also follows from \eqref{eq:modelCondition} that 
\begin{equation} \label{eq:p0jEq}
p(0, j ) = c p(j, 0).    
\end{equation}

We claim that the unique invariant measure of $V(t)$ (up to multiplication by a constant coefficient) is given by $\invMeasure{l} = c$ for all $l \neq 0$ and $\invMeasure{0} = 1$. We check this by showing  $\invMeasure{\cdot}$ satisfies the detailed balance equation,
\begin{equation}\label{eq:detailedBalance}
\invMeasure{i} p(i,j) = \invMeasure{j} p(j,i)
\end{equation} 
for all $i, j \in \Z$. This also shows that the walk $V(t)$ is reversible with respect to $\invMeasure{\cdot}$.

The detailed balance condition is clearly true for the case $i = j = 0$, and for $i, j \neq 0$ it follows from \eqref{eq:pijEquality}. We now consider the case when $i = 0$ and $j \neq 0$. Substituting $\invMeasure{j} = c$ and $\invMeasure{0} = 1$ into \eqref{eq:detailedBalance}, we obtain
\begin{equation*}
p(0, j) = c p(j, 0),
\end{equation*} 
which is true by \eqref{eq:p0jEq}. Therefore, our claim that $\invMeasure{l} = c$ for $l\neq 0$ and $\invMeasure{0} = 1$ is justified.

It follows from \eqref{eq:absMeasureDef} that 
\begin{equation}\label{eq:absMeasureFormula}
    \absMeasure{l} = 
    \begin{cases}
        1
 &\text{if $i=0$} \\
  2c &\text{if $i \neq 0$}  \end{cases}
\end{equation}
as desired.

\emph{Simplification of the expectation value:} We now simplify $\expTotal{\diffRandomWalk{t+1} - \diffRandomWalk{t} \mid \diffRandomWalk{t} = l}$ to show it is given by \eqref{eq:expSimplified}. For $l=0$, the two walkers $R^1$ and $R^2$ are both at the same site; therefore, they use the same jump rate such that
\begin{equation*}
    \expTotal{\diffRandomWalk{t+1} - \diffRandomWalk{t} \mid \diffRandomWalk{t} = 0} = \sum_{i, j \in \Z} |i - j| \expAnnealed{\xi(i) \xi(j)},
\end{equation*}
which agrees with \eqref{eq:expSimplified}.

For $l > 0$, the two walkers $R^1$ and $R^2$ are at different sites; therefore, they use independent jump rates such that
\begin{equation}\label{eq:expGeneral}
    \expTotal{\diffRandomWalk{t+1} - \diffRandomWalk{t} \mid \diffRandomWalk{t} = l} = \sum_{i,j \in \Z} (|l+i-j| - l) \meanOmega{i}\meanOmega{j}. 
\end{equation}
We now break this sum up into two parts: when $|i-j| \leq l$ and $|i-j| > l$. For $|i-j| \leq l$, we simplify
$$
|l+i-j|  - l = \begin{cases} i-j \quad \text{if } i \geq j \\ j-i \quad  \text{if } i<j,\end{cases}.
$$
Therefore, 
\begin{align*}
    \sum_{|i-j| \leq l} (|l+i-j| - l) \meanOmega{i}\meanOmega{j} &= \sum_{i < j} (j-i) \meanOmega{i}\meanOmega{j} + \sum_{i \geq j} (i -j) \meanOmega{i} \meanOmega{j} \\
    &= 0
\end{align*}
after swapping the indices of the second sum. 

For $|i-j| > l$, we find 
$$
|l+i-j|  - l = \begin{cases} i-j  \quad \text{if } i \geq j \\ j-i -2l \quad  \text{if } i <j\end{cases}.
$$
Substituting this into \eqref{eq:expGeneral}, we are left with 
\begin{align*}
    \expTotal{\diffRandomWalk{t+1} - \diffRandomWalk{t} \mid \diffRandomWalk{t} = l} &= \sum_{|i-j|>  l} (|l+ i -j| - l)\meanOmega{i}\meanOmega{j} \\
    &= \sum_{i-j>  l} (|i-j|)\meanOmega{i}\meanOmega{j} + \sum_{j-i>  l} (|i-j| - 2l)\meanOmega{i}\meanOmega{j}  \\
    &= \sum_{|i-j|>  l} (|i -j| - l)\meanOmega{i}\meanOmega{j}.
\end{align*}
Thus, $\expTotal{\diffRandomWalk{t+1} - \diffRandomWalk{t} \mid \diffRandomWalk{t} = l}$ is given by \eqref{eq:expSimplified}.

\emph{Computing $\extCoef$: } We now compute $\extCoef$ using our formula for $\absMeasure{l}$ and $\expTotal{\diffRandomWalk{t+1} - \diffRandomWalk{t} \mid \diffRandomWalk{t} = l}$. Substituting \eqref{eq:absMeasureFormula} and \eqref{eq:expSimplified}, we find
\begin{align}
    \sum_{l=0}^\infty \expTotal{\diffRandomWalk{t+1} - \diffRandomWalk{t} \mid \diffRandomWalk{t} = l} \absMeasure{l} &= c \sum_{i, j \in \Z} |i-j| \meanOmega{i} \meanOmega{j} + 2 c \sum_{l=1}^\infty \sum_{|i-j|>l} (|i-j|-l) \meanOmega{i}\meanOmega{j} \\
    &= c \sum_{l=0}^{\infty} \sum_{|i-j|>l}(2-\mathbbm{1}_{l=0})(|i-j|-l) \meanOmega{i} \meanOmega{j} \label{eq:sumOverl}
\end{align}
after combining the two sums. Now we break up the second sum over $i$ and $j$ such that 
\begin{align*}
     \sum_{l=0}^\infty & \expTotal{\diffRandomWalk{t+1} - \diffRandomWalk{t} \mid \diffRandomWalk{t} = l} \absMeasure{l} \\ 
     &= c \sum_{l=0}^{\infty} \sum_{j \in \Z} \left[ \sum_{i=j+l+1}^{\infty} (2 - \mathbbm{1}_{l=0})(i-j-l) \meanOmega{i} \meanOmega{j} + \sum_{i=-\infty}^{j-l-1} ( 2 - \mathbbm{1}_{l=0}) (j-i-l)\meanOmega{i} \meanOmega{j}\right] \\
    &= c \sum_{j \in \Z} \left[ \sum_{i=j+1}^\infty \sum_{l=0}^{i-j-1} (2-\mathbbm{1}_{l=0}) ( i-j -l) \meanOmega{i} \meanOmega{j} + \sum_{i=-\infty}^{j-1} \sum_{l=0}^{j-i-1} (2 - \mathbbm{1}_{l=0})(j-i-l) \meanOmega{i} \meanOmega{j} \right] ,
\end{align*}
where in the second step, we switch the order of the sums (which is justified by our assumption that $\xi$ is finite range).  Notice that since $\meanOmega{i}$ does not depend on $l$ we can now evaluate the sums over $l$. This yields 
\begin{align*}
     \sum_{l=0}^\infty \expTotal{\diffRandomWalk{t+1} - \diffRandomWalk{t} \mid \diffRandomWalk{t} = l} \absMeasure{l}  &= c \sum_{j\in\Z} \left[ \sum_{i=j+1}^{\infty} (i-j)^2 \meanOmega{i} \meanOmega{j} + \sum_{i=-\infty}^{j-1} (i-j)^2 \meanOmega{i} \meanOmega{j} \right]  \\
    &= c \sum_{i, j \in \Z} (i-j)^2 \meanOmega{i} \meanOmega{j} \\ 
    &= c \sum_{i, j \in \Z} (i^2 +j^2) \meanOmega{i} \meanOmega{j} \\
    &= 2 c \sum_{i\in\Z} i^2 \meanOmega{i}
\end{align*}
after using $\sum_{i\in\Z} \meanOmega{i} i = 0$. Thus, we find 
\begin{equation}\label{eq:generalLambdaCase}
    \extCoef = \frac{\driftVar}{ \sum_{l=0}^\infty \expTotal{\diffRandomWalk{t+1} - \diffRandomWalk{t} \mid \diffRandomWalk{t} = l} \absMeasure{l} } = \frac{1-c}{2c}. 
\end{equation}
We now match this to \eqref{eq:lambdaSimplified}. We recall that $D = \frac{1}{2} \sum_{i \in \Z} i^2 \meanOmega{i}$ and that $\dExt = \frac{1}{2} \driftVar$ where $\driftVar$ is given in \eqref{eq:generalDriftVar}. Therefore,
\begin{equation}
    \frac{\dExt}{2 ( D -\dExt)} = \frac{1-c}{2c} = \extCoef.
\end{equation}

In the next several sections, we use the above simplification to derive $\extCoef$ for several explicit examples. See Table \ref{tab:examples} for a summary.

{\renewcommand{\arraystretch}{3.5}
\begin{table}
    \centering
    \begin{tabular}{|l||c|c|c|c|} \hline 
          &Dirichlet Distribution&  Normalized i.i.d. Distribution &  Random Delta Distribution& Nearest Neighbor Distribution\\ \hline 
          $c$ & $\displaystyle \frac{\alpha}{\alpha + 1}$& $c$  & $\displaystyle\frac{2k+1}{4k}$  &  $2(1-2\mathbb{E}[\omega^2])$  \\ \hline 
          $D$ & $\displaystyle \frac{1}{2 \alpha} \sum_{i =-k}^k \alpha_{i} i^2 $ & $\displaystyle \frac{1}{6} k (k+1)$ & $\displaystyle \frac{1}{6}k (k+1)$ & $\displaystyle\frac{1}{2}$ \\ \hline 
          $\dExt$ & $\displaystyle \frac{1}{2\alpha (\alpha + 1)} \sum_{i = -k}^k \alpha_{i} i^2$ & $\displaystyle \frac{1}{6} k (k+1)(1-c)$ & $\displaystyle \frac{1}{24} (k+1)(2k-1)$ &$2 \mathbb{E}[\omega^2] - \frac{1}{2} $ \\ \hline 
          $\extCoef$ & $\displaystyle \frac{1}{2 \alpha}$  & $\displaystyle\frac{1-c}{2c}$  & $\displaystyle\frac{2k-1}{2(2k+1)}$ & $ \displaystyle \frac{4 \mathbb{E}[\omega^2] -1}{2(1-2\mathbb{E}[\omega^2])}$\\ \hline
    \end{tabular}
    \caption{A summary of the relevant coefficients for the examples in Sections \ref{sec:dirichlet}, \ref{sec:iid}, \ref{sec:delta}, and \ref{sec:NearestNeighbor}.}
    \label{tab:examples}
\end{table}}

\subsection{Dirichlet distribution}\label{sec:dirichlet}

We recall our definition of the Dirichlet distribution from Section \ref{sec:DistributionDefinitions}. The Dirichlet distribution for $\nu$ is specified by a choice of $k\in \mathbb{Z}_{\geq 1}$ and $\vec{\alpha} = (\alpha_{-k},\ldots,\alpha_{k})\in \mathbb{R}_{>0}^{2k+1}$. It assigns probability density proportional to  $\prod_{j=-k}^{k} \xi(j)^{\alpha_j}$ to each vector $\xi=(\xi(-k),\ldots, \xi(k)) \in \mathbb{R}_{ \geq 0}^k$ satisfying $\sum_{j=-k}^{k} \xi(j) = 1.$ Then, we set $\xi(i) = 0$ for $i \notin [-k, k]$ so that $\xi$ is defined on $\Z$. 

We have 
\begin{align*}\label{eq:DirichletMean}
\meanOmega{i} &= \begin{cases}
        \frac{\alpha_{i}}{\alpha} & \text{for } -k \leq i \leq k \\
        0  & \text{else},
    \end{cases}
\end{align*}
and for $i, j \in \mathbb{Z}$, $i \neq j$:
\begin{align*}
    \expAnnealed{\xi(i)\xi(j)} &= \frac{ \alpha_{i}\alpha_{j}}{\alpha(\alpha+1)} =  \frac{\alpha }{\alpha+1}\meanOmega{i} \meanOmega{j}.
\end{align*}
It follows that \eqref{eq:modelCondition} and \eqref{eq:irreducibility} hold for the Dirichlet distribution with $c = \frac{\alpha}{\alpha + 1}$ so that the Dirichlet distribution falls into the general class of models considered above. Combining \eqref{eq:dExt} and \eqref{eq:generalDriftVar}, we see that
\begin{equation*}
    \dExt = \frac{1}{2\alpha (\alpha + 1)} \sum_{i = -k}^k \alpha_{i} i^2.
\end{equation*}
We calculate the diffusion coefficient from its definition in \eqref{eq:edc},
\begin{equation*}
    D = \frac{1}{2 \alpha} \sum_{i = -k}^k \alpha_{i} i^2. 
\end{equation*}
Finally, we see from \eqref{eq:generalLambdaCase} that 
\begin{equation}
\extCoef = \frac{1}{2\alpha}.
\end{equation}

\subsection{Independent and Identically Distributed Random Variables}\label{sec:iid}

We recall our definition of the normalized i.i.d. distribution from Section \ref{sec:DistributionDefinitions}. The normalized i.i.d. distribution for $\nu$ is specified by a choice of $k\in \mathbb{Z}_{\geq 1}$ and a choice of probability distribution on $\mathbb{R}_{\geq 0}$. Let $X_{-k},\ldots, X_{k}$ be chosen independently according to the specified probability distribution, and then define $\xi(i) = X_i (\sum_{j=-k}^{k} X_j)^{-1}$ for $i \in [-k, k]$ and $\xi(i) = 0$ otherwise.

We have 
\begin{align*}
\meanOmega{i} &= \begin{cases}
        \frac{1}{2k+1} & \text{for } -k \leq i \leq k \\
        0  & \text{else},
    \end{cases}
\end{align*}
and for $i, j \in \mathbb{Z}$, $i \neq j$:
\begin{align*}
    \expAnnealed{\xi(i)\xi(j)} &= \frac{c}{(2k+1)^2}
\end{align*}
for some constant $c \in (0,1)$. Therefore, \eqref{eq:modelCondition} and \eqref{eq:irreducibility} are satisfied. Combining \eqref{eq:dExt} and \eqref{eq:generalDriftVar}, we see that
\begin{equation*}
    \dExt = \frac{1-c}{2(2k+1)} \sum_{i  = -k}^k i^2 = \frac{1}{6} k (k+1)(1-c)
\end{equation*}
We calculate the diffusion coefficient from its definition in \eqref{eq:edc},
\begin{equation*}
    D = \frac{1}{2 (2k+1)} \sum_{i = - k}^k i^2 = \frac{1}{6} k (k+1). 
\end{equation*}
Finally, we see from \eqref{eq:generalLambdaCase} that 
\begin{equation}
     \extCoef = \frac{1-c}{2c}.
\end{equation}




\subsection{Random Delta Distribution}\label{sec:delta}

We recall our definition of the random delta distribution from Section \ref{sec:DistributionDefinitions}. The \emph{random delta} distribution for $\nu$ is specified by $k \in \Z_{\geq 1}$. Two numbers $X_1,X_2$ are drawn uniformly without replacement from $\{-k,\ldots, k\}$. We set $\xi(X_1)=\xi(X_2)=1/2$ and $\xi(i)=0$ for all $i \neq X_1, X_2$.

We have 
\begin{align*}
\meanOmega{i} &= \begin{cases}
        \frac{1}{2k + 1} & \text{for } -k \leq i \leq k \\
        0  & \text{else},
    \end{cases}
\end{align*}
and for $i, j \in \mathbb{Z}$, $i \neq j$:
\begin{align*}
    \expAnnealed{\xi(i) \xi(j)} &= \frac{1}{4k(2k+1)}.
\end{align*}
It follows that \eqref{eq:modelCondition} and \eqref{eq:irreducibility} are satisfied with $c = \frac{2k+1}{4k}$. Combining \eqref{eq:dExt} and \eqref{eq:generalDriftVar}, we see that
\begin{align*}
    \dExt = \frac{2k-1}{8k (2k+1)} \sum_{i= - k}^{k} i^2 = \frac{1}{24} (k+1) (2k-1).
\end{align*}
We calculate the diffusion coefficient from its definition in \eqref{eq:edc},
\begin{equation*}
    D = \frac{1}{6} k (k+1). 
\end{equation*}
Finally, we see from \eqref{eq:generalLambdaCase} that
\begin{equation}
    \extCoef = \frac{2k-1}{2(2k+1)}.
\end{equation}


\subsection{Nearest Neighbor Distribution}\label{sec:NearestNeighbor} 

We define the nearest neighbor distribution as follows. Let $\omega$ be any random variable taking values in the interval $[0,1]$ such that $\mathbb{E}[\omega] = \frac{1}{2}$. We set $\xi(1) = \omega$, $\xi(-1) = 1 - \omega$ and all other $\xi(i)=0$. 

We have

\begin{align*}
     \meanOmega{1} &=\meanOmega{-1}= \frac{1}{2}
\end{align*}
and
\begin{align*}
\expAnnealed{\xi(1) \xi(-1)} &=\frac{1}{2}-\mathbb{E}[\omega^2].  
\end{align*}
It follows that \eqref{eq:modelCondition} is satisfied with $c = 2(1-2\mathbb{E}[\omega^2])$; however, the walk $V(t)$ is no longer irreducible. In fact, when started from $0$, it remains restricted to the sublattice $2\mathbb{Z}$. Therefore, $\invMeasure{\cdot} = 0$ for all $l \notin 2\mathbb{Z}$, $\invMeasure{0} = 1$ and  $\invMeasure{l} = c$ for $l \in 2\mathbb{Z}$. This slightly changes the analysis performed above since up until now, we were assuming that $\invMeasure{l} = c$ for all $l \neq 0$. In particular, \eqref{eq:sumOverl} is replaced by 

\begin{align}
    \sum_{l=0}^\infty \expTotal{\diffRandomWalk{t+1} - \diffRandomWalk{t} \mid \diffRandomWalk{t} = l} \absMeasure{l} &=c \sum_{l \in 2\mathbb{Z}_{\geq 0}}\sum_{|i-j|>l}(2-\mathbbm{1}_{l=0})(|i-j|-l) \meanOmega{i} \meanOmega{j}.
\end{align}
In other words, we are only summing over even integers $l$. This can be further simplified in the same manner as above such that
$$
    \sum_{l=0}^\infty \expTotal{\diffRandomWalk{t+1} - \diffRandomWalk{t} \mid \diffRandomWalk{t} = l} \absMeasure{l} = c \sum_{i\in\Z} i^2 \meanOmega{i}= c.
$$
Note that this is simply half of what we obtained when summing over all integers instead of just even ones. We can also see this more directly by noticing that $\expTotal{\diffRandomWalk{t+1} - \diffRandomWalk{t} \mid \diffRandomWalk{t} = l} = 0$ for $l \geq 2$. Therefore, 

$$
    \sum_{l=0}^{\infty}\invMeasure{l} \expAnnealed{\diffRandomWalk{t+1} - \diffRandomWalk{t} \mid \diffRandomWalk{t} = l} = \expAnnealed{\diffRandomWalk{t+1} - \diffRandomWalk{t} \mid \diffRandomWalk{t} = 0} 
    =  2(1-2\mathbb{E}[\omega^2])
     =c.
$$

We compute $\driftVar = 4 \mathbb{E}[\omega^2] - 1 = 1 - c$ so that 
\begin{equation}
    \extCoef = \frac{\driftVar}{ \sum_{l=0}^\infty \expTotal{\diffRandomWalk{t+1} - \diffRandomWalk{t} \mid \diffRandomWalk{t} = l} \absMeasure{l} } = \frac{4 \mathbb{E}[\omega^2] -1}{2(1-2\mathbb{E}[\omega^2])} = \frac{1-c}{c}.
\end{equation} 
This matches with the results in \cite{dasKPZEquationLimit2023c}. Since $D = 1/2$, we have that 
\begin{equation}
    \extCoef = \frac{\driftVar}{(2D - \driftVar)},
\end{equation} 
which is twice that of the irreducible cases. Similar extensions are possible if $V(t)$ is restricted to other sublattices, but we do not write out the details here.

\end{document}